\documentclass[fleqn,usenatbib]{mnras}
\pdfoutput=1

\usepackage[T1]{fontenc}
\usepackage{ae,aecompl}
\usepackage{longtable}
\usepackage{supertabular}

\usepackage{graphicx}	
\usepackage{amsmath}	
\usepackage{amssymb}	
\usepackage{newtxtext,newtxmath}

\title[Cluster candidates in the \textit{AKARI} NEP field by HSC]{Optically-detected galaxy cluster candidates in the \textit{AKARI} North Ecliptic Pole field based on photometric redshift from Subaru Hyper Suprime-Cam}

\author[T.-C. Huang et al.]{
Ting-Chi Huang \href{https://orcid.org/0000-0001-7200-8157}{\includegraphics[scale=0.1]{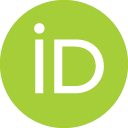}},$^{1,2}$\thanks{E-mail: kevintch@ir.isas.jaxa.jp}
Hideo Matsuhara,$^{1,2}$
Tomotsugu Goto \href{https://orcid.org/0000-0002-6821-8669}{\includegraphics[scale=0.1]{ORCIDiD_icon128x128.png}},$^{3}$
\newauthor
Daryl Joe D. Santos \href{https://orcid.org/0000-0002-5687-0609}{\includegraphics[scale=0.1]{ORCIDiD_icon128x128.png}},$^{3}$
Simon C.-C. Ho \href{https://orcid.org/0000-0002-8560-3497}{\includegraphics[scale=0.1]{ORCIDiD_icon128x128.png}},$^{3}$
Seong Jin Kim \href{https://orcid.org/0000-0001-9970-8145}{\includegraphics[scale=0.1]{ORCIDiD_icon128x128.png}},$^{3}$
\newauthor
Tetsuya Hashimoto \href{https://orcid.org/0000-0001-7228-1428}{\includegraphics[scale=0.1]{ORCIDiD_icon128x128.png}},$^{3,4}$
Hiroyuki Ikeda,$^{5,6}$
Nagisa Oi \href{https://orcid.org/0000-0002-4686-4985}{\includegraphics[scale=0.1]{ORCIDiD_icon128x128.png}},$^{7}$
\newauthor
Matthew A. Malkan \href{https://orcid.org/0000-0001-6919-1237}{\includegraphics[scale=0.1]{ORCIDiD_icon128x128.png}},$^{8}$
William J. Pearson \href{https://orcid.org/0000-0002-7300-2213}{\includegraphics[scale=0.1]{ORCIDiD_icon128x128.png}},$^{9}$
Agnieszka Pollo, $^{9,10}$
\newauthor
Stephen Serjeant,$^{11}$
Hyunjin Shim \href{https://orcid.org/0000-0002-4179-2628}{\includegraphics[scale=0.1]{ORCIDiD_icon128x128.png}},$^{12}$
Takamitsu Miyaji \href{https://orcid.org/0000-0002-7562-485X}{\includegraphics[scale=0.1]{ORCIDiD_icon128x128.png}},$^{13,14}$
\newauthor
Ho Seong Hwang \href{https://orcid.org/0000-0003-3428-7612}{\includegraphics[scale=0.1]{ORCIDiD_icon128x128.png}},$^{15}$
Anna Durkalec \href{https://orcid.org/0000-0002-3818-8315}{\includegraphics[scale=0.1]{ORCIDiD_icon128x128.png}},$^{9}$
Artem Poliszczuk,$^{9}$
\newauthor
Thomas R.~Greve,\href{https://orcid.org/0000-0002-2554-1837}{\includegraphics[scale=0.1]{ORCIDiD_icon128x128.png}},$^{16,17,18}$
Chris Pearson,$^{11,19,20}$
Yoshiki Toba \href{https://orcid.org/0000-0002-3531-7863}{\includegraphics[scale=0.1]{ORCIDiD_icon128x128.png}},$^{21,22,23}$
\newauthor
Dongseob Lee,$^{12}$
Helen K. Kim,$^{8}$
Sune Toft,$^{17,24}$
Woong-Seob Jeong,$^{25}$
\newauthor
and Umi Enokidani,$^{1,2}$
\\
Author Affiliations are listed in Appendix B}

\date{Accepted XXX. Received YYY; in original form ZZZ}

\pubyear{2021}

\begin{document}
\label{firstpage}
\pagerange{\pageref{firstpage}--\pageref{lastpage}}
\maketitle

\begin{abstract}
Galaxy clusters provide an excellent probe in various research fields in astrophysics and cosmology. However, the number of galaxy clusters detected so far in the \textit{AKARI} North Ecliptic Pole (NEP) field is limited. In this work, we provide galaxy cluster candidates in the \textit{AKARI} NEP field with the minimum requisites based only on coordinates and photometric redshift (photo-$z$) of galaxies. We used galaxies detected in 5 optical bands ($g$, $r$, $i$, $z$, and $Y$) by the Subaru Hyper Suprime-Cam (HSC), assisted with $u$-band from Canada-France-Hawaii Telescope (CFHT) MegaPrime/MegaCam, and IRAC1 and IRAC2 bands from the \textit{Spitzer} space telescope for photo-$z$ estimation. We calculated the local density around every galaxy using the 10$^{th}$-nearest neighbourhood. Cluster candidates were determined by applying the friends-of-friends algorithm to over-densities. 88 cluster candidates containing 4390 member galaxies below redshift 1.1 in 5.4 deg$^2$ have been detected. The reliability of our method was examined through false detection tests, redshift uncertainty tests, and applications on the COSMOS data, giving false detection rates of 0.01 to 0.05 and recovery rate of 0.9 at high richness. 3 X-ray clusters previously observed by \textit{ROSAT} and \textit{Chandra} were recovered. The cluster galaxies show higher stellar mass and lower star formation rate (SFR) compared to the field galaxies in two-sample Z-tests. These cluster candidates are useful for environmental studies of galaxy evolution and future astronomical surveys in the NEP, where \textit{AKARI} has performed unique 9-band mid-infrared photometry for tens of thousands of galaxies.
\end{abstract}




\section{Introduction}
\label{intro}
Cluster of galaxies are very important astronomical objects, for their wide usages in the large varieties of research fields in both astrophysics and cosmology. For example, because of their high density, clusters of galaxies are good laboratories for studying the environmental effect on galaxy evolution \citep[e.g.,][]{Goto2003,Park2009,Vulcani2010,Hwang2012}. In addition, clusters of galaxies, as massively bound systems in the large-scale structure, can provide the measurement of the standard ruler in the baryon acoustic oscillations for cosmological studies \citep[e.g.,][]{Hong2012,Hong2016}. Clusters of galaxies are also frequently used as gravitational lenses to observe faint or high-redshift galaxies \citep[e.g.,][]{Coe2013}. 

There are many approaches to detect galaxy clusters using their various characteristics in different wavelength. The intracluster medium consists of hot gas at a temperature of more than $10^{8}$ K due to inefficient cooling by the bremsstrahlung process, emitting radiation in X-ray. Therefore, identifying galaxy clusters from X-ray surveys is relatively straightforward \citep[e.g.,][]{Boehringer2001,Takey2011}. Meanwhile, the high-energy electrons interact with the cosmic microwave background (CMB) photons through the inverse Compton scattering, and distort the CMB spectrum, known as the Sunyaev-Zeldovich (SZ) effect \citep{SZ1980}. Thus, galaxy clusters are also frequently detected at millimeter wavelength by either ground-based \citep[e.g.,][]{Staniszewski2009,Bleem2015,Hilton2020} or space-based \citep[e.g.,][]{Planck2014,Planck2016} telescopes. Nowadays, more and more cluster finding algorithms using optical photometric or spectroscopic data of galaxies have been developed. These methods generally utilise the colour of the cluster red sequence \citep[e.g.,][]{Muzzin2013,Rykoff2014,Oguri2018}, the density enhancement associated with the galaxy redshift \citep[e.g.,][]{Wen2012,Tempel2014,Hung2020}, or both \citep[e.g.,][]{Goto2002,Goto2008}. All of these approaches have their advantages and disadvantages. Using the optical photometric data allows us to find as many galaxy clusters as possible, since they are generally observed by large-area deep surveys, providing a numerous sample of galaxies. In this work, we used optical photometric data from HSC to search for clusters in the \textit{AKARI} NEP field. 

\textit{AKARI} is an infrared space telescope launched by the Japan Aerospace Exploration Agency (JAXA) in 2006 \citep{Murakami2007}. \textit{AKARI} carried out a 5.4-deg$^{2}$ survey at NEP \citep[][]{Kim2012} using its Infrared Camera \citep[IRC;][]{Onaka2007}, which is equipped with 9 filters in mid-infrared ($N2, N3, N4, S7, S9W, S11, L15, L18W,$ and $L24$). The number in the name of each filter corresponds to its effective wavelength. \textit{AKARI}/IRC's 9 filters provide us with a unique 9-band photometry continuously covering the whole mid-infrared range \citep{Matsuhara2006}, which makes \textit{AKARI} superior to other infrared telescopes such as \textit{Spitzer} or \textit{WISE} in terms of the wavelength coverage. With this advantage, we are able to analyse the mid-infrared features of low-redshift galaxies more accurately: for example, the polycyclic aromatic hydrocarbons (PAH) emission features of star-forming galaxies \citep[e.g.,][]{Murata2014,Kim2019} or the warm dust emissions from active galactic nuclei \citep[AGN; e.g.,][]{Huang2017,Toba2020,Wang2020}. Moreover, the \textit{AKARI} NEP data also show great performance in AGN selection based on machine learning methods \citep[e.g.][]{Poliszczuk2019,Chen2021,Poliszczuk2021}. The \textit{AKARI} NEP field has been extensively observed in multi-wavelength \citep[e.g.][]{Kim2012,Kim2021,Oi2014}, and many studies in galaxy evolution are expected to be performed with these fruitful data in the future: for example, environmental effects on galaxy evolution and AGN activities (Santos et al. submitted). 

Although the \textit{AKARI} NEP data are very useful as previously described, so far, only a small number of galaxy clusters have been found in the \textit{AKARI} NEP field. \cite{Henry2006} reported the \textit{ROSAT} NEP X-ray survey catalogue, and categorised 62 clusters of galaxies. Only 7 of them are located in the \textit{AKARI} NEP field. \cite{Goto2008} found 16 galaxy cluster candidates in the \textit{AKARI} NEP deep field \citep{Murata2013} in 0.5 deg$^2$ at redshift $0.9 < z < 1.7$ based on a colour-cut method. \cite{Ko2012} reported the discovery of a supercluster at $z=0.087$ in the NEP. The number of galaxy clusters identified so far is not sufficient for statistical analysis. Also, the clusters' redshifts are not continuously distributed in the whole redshift range, so that the study of the evolution of the galaxies in clusters (cluster galaxies hereafter) has still been limited. Hence, extending the search of galaxy clusters to the entire \textit{AKARI} NEP field in a large redshift range is necessary, and also beneficial to any further studies related to galaxy clusters such as \textit{Euclid} \citep{Laureijs2011}. Fortunately, an optical survey was conducted on the whole \textit{AKARI} NEP field \citep{Goto2017} by Subaru/HSC \citep{Miyazaki2012}, providing deep optical data which can be utilised to search for galaxy clusters.

There are two goals in this study. The first is (1) to provide a catalogue of potential cluster candidates for studying environmental effects in the \textit{AKARI} NEP field with as few requisites as possible. The reason is that the number of clusters in the \textit{AKARI} NEP field is limited, so it is undesirable to remove sources due to any assumption or selection, which may reduce the number of potential cluster candidates. By doing so, we can also avoid imposing too many biases on them, providing a more complete sample of clusters for environmental studies. The second is (2) to provide a useful method for finding galaxy clusters, which can be applied to any optical photometric survey. Despite the substantial number of cluster finding methods available today, there are never too many methods for cluster finding. Each method has different advantages and disadvantages that are suitable for different situations. Our presented method has the advantage that it is simple and quick, allowing a straightforward operation.

In this paper, we describe our data and method in Section~\ref{DandA}. Our results including galaxy colour, richness, a catalogue and optical images of our cluster candidates, and reliability tests are presented in Section~\ref{results}. In Section~\ref{discussion}, we discuss a comparison with X-ray clusters, a comparison with COSMOS clusters, properties of cluster galaxies, the influence of spectroscopic data, and future prospects. A summary is given in Section~\ref{summary}. Throughout this paper, the cosmology is assumed as the following: $H_{0}=70$ km s$^{-1}$ Mpc$^{-1}$, $\Omega_{m}=0.3$, and $\Omega_{\Lambda}=0.7$. We use the AB magnitude system unless otherwise mentioned.

\section{Data and Analysis}
\label{DandA}
The data and the sample selection are described in Section~\ref{sec:data}, while the cluster finding methods including the photo-$z$, density calculation and the friends-of-friends algorithm are described in Section~\ref{analysis}.

\subsection{Data and sample selection}
\label{sec:data} 
\begin{figure*}
	\includegraphics[width=1.5\columnwidth]{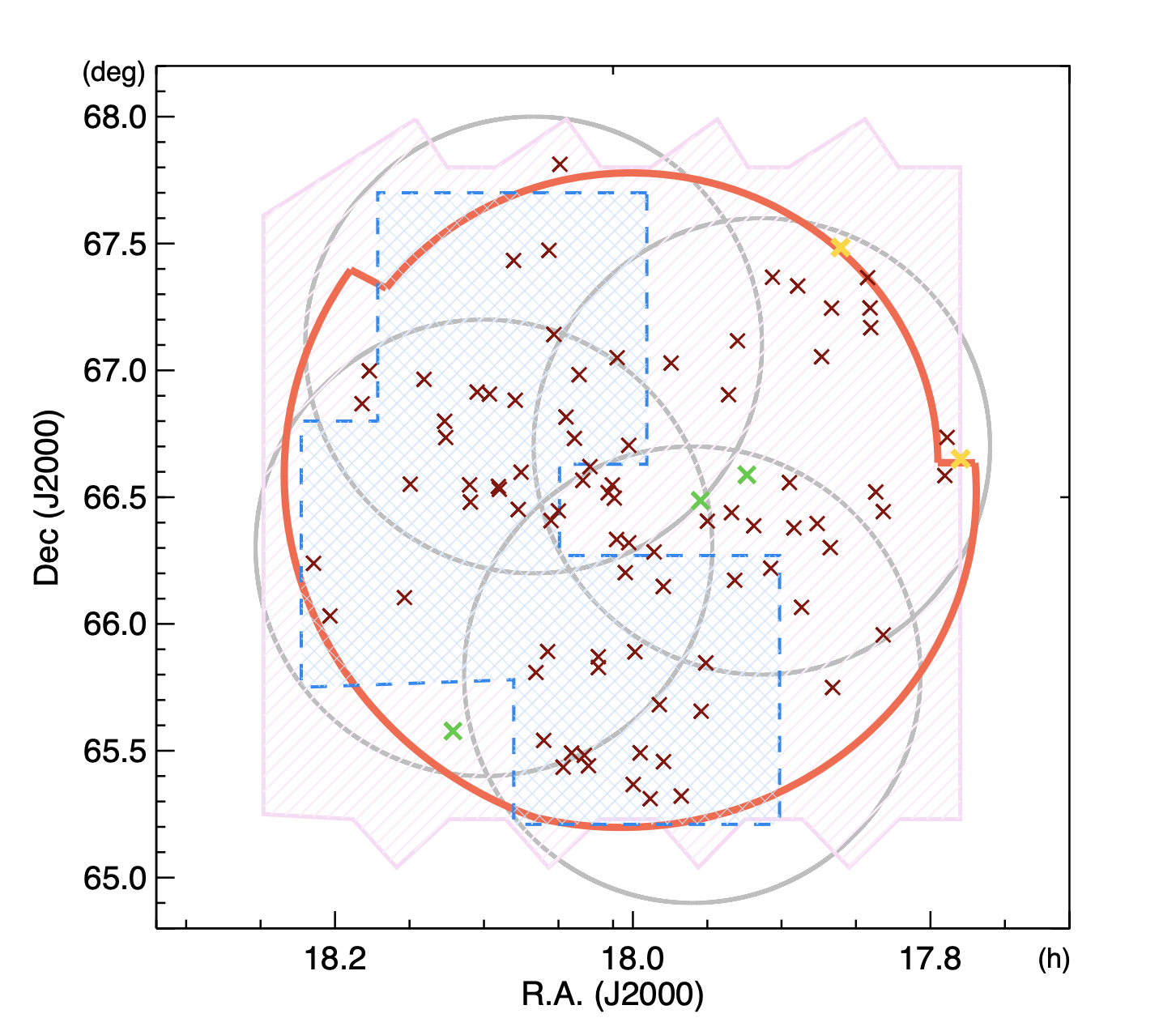}
\caption{ The sky map showing the coverages of the HSC observations (grey), the CFHT observations (blue), and the 
\textit{Spitzer} observations (pink) in the \textit{AKARI} NEP field (red). The cluster candidates are plotted as crosses, while the candidates matched with X-ray clusters and the candidates near the edge of the HSC survey are plotted in green and orange, respectively. }
    \label{fig:sky}
\end{figure*}

We used the source catalogue from the HSC survey \citep{Goto2017,Oi2021} in the \textit{AKARI} NEP field to detect galaxy clusters using the friends-of-friends algorithm \citep{HG1982}, which requires only the coordinates and the redshifts of galaxies. For the redshifts, we used the following photometric dataset to estimate the photo-$z$: (1) HSC $g$, $r$, $i$, $z$, and $y$ bands \citep{Oi2021}, (2) CFHT MegaPrime/MegaCam $u$ band \citep{Huang2020}, and (3) \textit{Spitzer} IRAC 3.6-$\mu$m, and 4.5-$\mu$m bands \citep{Nayyeri2018}. The data (2) and (3) are included wherever available. Fig.~\ref{fig:sky} shows the \textit{AKARI} NEP field (red), the HSC observation coverage (grey), the CFHT observation coverage (blue), and the \textit{Spitzer} observation coverage (pink). The crosses represent the positions of cluster candidates (to be discussed in Section~\ref{results}).

\subsubsection{Data}
\label{data}
The Subaru HSC data used in this work are described in \cite{Oi2021} and \cite{Kim2021}, including the details of the observations and the data reduction. The total exposure times range from 5360 ($i$-band) to 27440 ($g$-band) seconds in each band. The 5$\sigma$ detection limits of magnitudes for $g, r, i, z,$ and $y$ bands are 28.6, 27.3, 26.7, 26.0, and 25.6 mag, respectively. There are more than 3 million sources in the original HSC catalogue, and 747673 sources have 5$\sigma$ detection in all the five HSC bands.

The details of the observations and the data reduction of the $u$-band data are described in \cite{Huang2020}. The CFHT $u$-band survey observed 3.6 deg$^2$ in the \textit{AKARI} NEP-Wide field and obtained a total integration time of 4520 to 13910 seconds. The $u$-band data have a 5$\sigma$ limiting magnitude of 25.4 mag for the whole \textit{AKARI} NEP-Wide field, while they reach 25.8 mag in a 1-deg$^{2}$ deeper area. 24 per cent of the HSC 5$\sigma$ sources are detected in the $u$-band data.

The \textit{Spitzer} IRAC 3.6-$\mu$m- and 4.5-$\mu$m-band data come from \cite{Nayyeri2018}. The \textit{Spitzer} observations surveyed a 7.04-deg$^{2}$ area at NEP, fully covering the \textit{AKARI} NEP field. The 5$\sigma$ depths of 3.6- and 4.5-$\mu$m bands are 21.9 and 22.5 mag, respectively. The fraction of the \textit{Spitzer} sources to the HSC 5$\sigma$ sources is 33 per cent. 

The data used for photo-$z$ calculation are summarised in Table ~\ref{tab:band}. In addition, we include 2146 spectroscopic sources at $z\leq1.1$ to check the photo-$z$ reliability and improve the cluster finding. The distribution of the spectroscopic redshift (spec-$z$) is plotted in blue in Fig.~\ref{fig:photoz_dist}. The spectroscopic data come from various telescopes: MMT/Hectospec and WIYN/Hydra \citep{Shim2013}, GTC/OSIRIS \citep{Diaz2017}, Subaru/FMOS \citep{Oi2018}, Keck/DEIMOS \citep{Takagi2010,Shogaki2018,Kim2018}, and \textit{AKARI}/IRC \citep{Ohyama2018}. The approach we used to match the above multi-wavelength data is described in \cite{Kim2021}.

\begin{table*}
	\centering
	\caption{The summary of the data used for the photo-$z$ calculation in this work.}
	\label{tab:band}
	\begin{tabular}{lcccr} 
		\hline
		Instrument & Area (deg$^{2}$) & Filter & Sensitivity (5$\sigma$, AB mag) & References\\
		\hline
		Subaru/HSC & 5.4 & $g$ & 28.6 & \cite{Oi2021}\\
		 & 5.4 & $r$ & 27.3 & \cite{Oi2021}\\
		 & 5.4 & $i$ & 26.7 & \cite{Oi2021}\\
		 & 5.4 & $z$ & 26.0 & \cite{Oi2021}\\
		 & 5.4 & $y$ & 25.6 & \cite{Oi2021}\\
		\hline 
		CFHT/MegaCam  & 3.6 & $u$ & 25.4 & \cite{Huang2020}\\
		\hline
		\textit{Spitzer}/IRAC & 7.0 & 3.6-$\mu$m & 21.9 & \cite{Nayyeri2018}\\
		 & 7.0 & 4.5-$\mu$m & 22.5 & \cite{Nayyeri2018}\\
		\hline
	\end{tabular}
\end{table*}

\subsubsection{Sample selection}
\label{sample}
We required the sources to have 5$\sigma$ detection in all the five HSC bands (i.e. sources fainter than 5$\sigma$ limiting magnitudes in any HSC band were rejected). An extended/point source classification using the HSC pipeline parameter \texttt{base$\_$ClassificationExtendedness$\_$value} was applied to exclude extended sources. A further star-galaxy separation was performed by the $\chi^2$ values from the template fitting (to be discussed in Section~\ref{photoz}) of galaxy ($\chi^2_{gal}$) and star ($\chi^2_{star}$) components: galaxies are selected if $\chi^2_{gal}$ < $\chi^2_{star}$. Since the photo-$z$ performance gets worse at higher redshift, we limited our sample up to redshift $z\leq1.1$. In total, the HSC sample contains 310577 sources (we call it selected sample hereafter). The CFHT $u$-band data and the \textit{Spitzer}/IRAC data are included with the selected sample to help the photo-z calculation if they are available.

\subsection{Analysis}
\label{analysis}

\subsubsection{Photometric redshift}
\label{photoz}
\begin{figure}
	\includegraphics[width=\columnwidth]{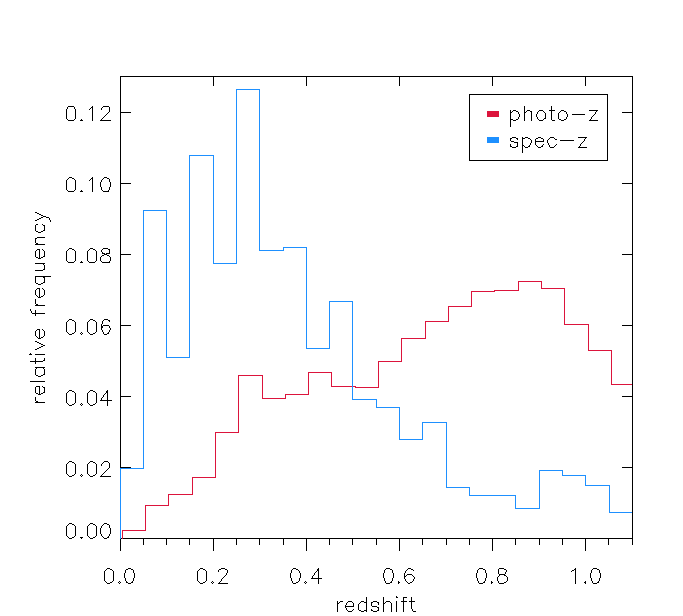}
\caption{ The redshift distributions of the spectroscopic sample of 2146 sources (blue; spec-$z$) and the selected sample of 310577 sources (red; photo-$z$) from redshift 0 to 1.1. The distribution is plotted in relative frequency, that is, the sum of all bins in each sample is equal to one. The bin size is 0.05. }
    \label{fig:photoz_dist}
\end{figure}

\begin{figure*}
	\includegraphics[width=\columnwidth]{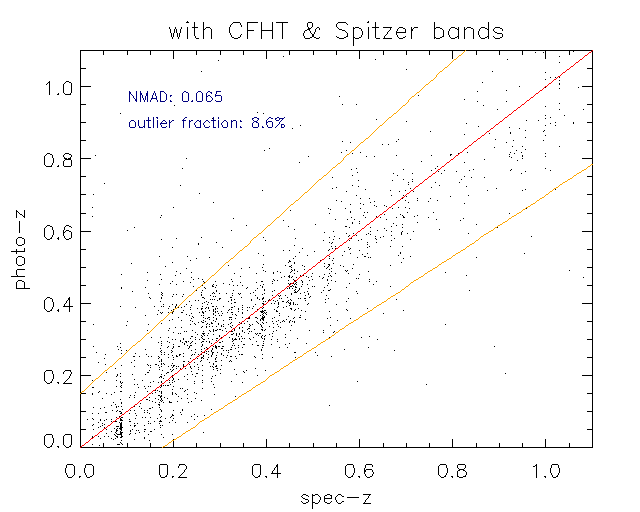}
	\includegraphics[width=\columnwidth]{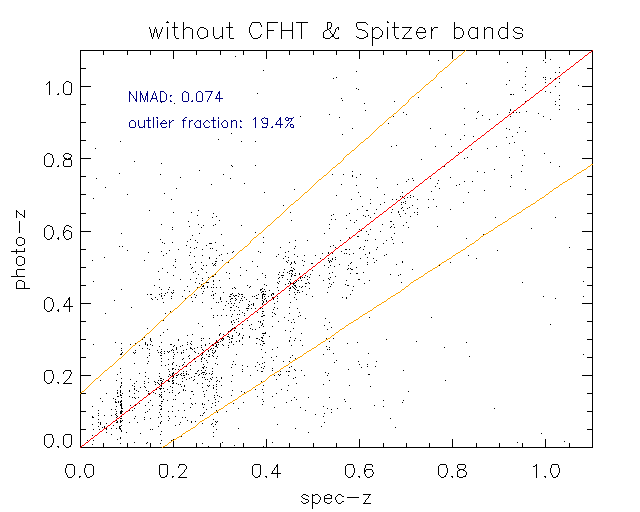}
    \caption{The comparison between photo-$z$ and spec-$z$ of the spec-$z$ sample. The photo-$z$ used in this work is plotted the left panel, while the photo-$z$ in the right panel is calculated without the CFHT and \textit{Spitzer} bands. The standard line (i.e. slope equals to one) is plotted in red. The orange lines show where $\frac{|z_{p}-z_{s}|}{1+z_{s}}=0.15$, which are the borders to define outliers. }
    \label{fig:photoz}
\end{figure*}

We used the Spectral Energy Distribution (SED) analysing code, \textsc{LePHARE} \citep{Arnouts1999,Ilbert2006}, to estimate the photo-$z$ (denoted as $z_{p}$ in equations) of our selected sample. The photo-$z$ distribution is shown in red in Fig.~\ref{fig:photoz_dist}. The method of the photo-$z$ calculation is the same as \cite{Ho2021}, but a smaller number of bands are used in this work. In total, 8 bands in maximum (Subaru/HSC $g$, $r$, $i$, $z$, $y$, CFHT/Megacam $u$, \textit{Spitzer}/IRAC 3.6-$\mu$m, 4.5-$\mu$m) were included in the SED fitting with stellar and galaxy templates. The stellar templates were combined from \cite{Bohlin1995}, \cite{Pickles1998}, and \cite{Chabrier2000}. They tested several templates for fitting galaxy components, and using the COSMOS templates \citep{Ilbert2009} gave them the most accurate performance in terms of the normalised median absolute deviation (NMAD, $\sigma_{z_{p}}$) and the catastrophic rate ($\eta$) for the subset of galaxies with spec-$z$. The NMAD is defined as $1.48 \times median(\frac{|z_{p}-z_{s}|}{1+z_{s}})$, where the $z_{s}$ denotes the spec-$z$. The catastrophic rate is defined by the fraction of the outliers in the spec-$z$ sample, while an object is regarded as an outlier if $\frac{|z_{p}-z_{s}|}{1+z_{s}} \ge 0.15$. In this work, our photo-$z$ performance are $\sigma_{z_{p}}$=0.065 (to be discussed in Section~\ref{lsd} and \ref{fof}) and $\eta$=8.6 per cent at $z_{p}\leq1.1$ ({the left panel of} Fig.~\ref{fig:photoz}). Throughout this work, if a galaxy has the spec-$z$ available then we use it instead. It is noteworthy that the photo-$z$ calculation improves remarkably if we make use of the CFHT and \textit{Spitzer} bands. The right panel of Fig.~\ref{fig:photoz} shows the performance of photo-$z$ calculated by using only five HSC bands, which has the NMAD of 0.074 and the outlier fraction of 19.4 per cent.

\subsubsection{Local surface density}
\label{lsd}
The 10$^{th}$-nearest neighbourhood method \citep[e.g.][]{Dressler1980,Miller2003} is used to calculate the local density around every galaxy. Due to the photo-$z$ uncertainty, the local density in this work is calculated in 2 dimensions within an individual redshift bin of every galaxy: $z \pm \sigma_{z_{p}} \times(1+z)$, where $\sigma_{z_{p}}=0.065$. We measured the angular separation $\theta$ between galaxy pairs as follows: 
\begin{equation}
    \theta_{1,2} = \cos^{-1}[\sin{\delta_{1}}\sin{\delta_{2}}+\cos{\delta_{1}}\cos{\delta_{2}}\cos{(\alpha_{1}-\alpha_{2})}], 
\end{equation}
where $\alpha \in [0,2\pi]$ and $\delta \in [\frac{-\pi}{2},\frac{\pi}{2}]$ are right ascension ($\alpha$) and declination ($\delta$), respectively.
Meanwhile, the 10$^{th}$-nearest local density of each object $\Sigma_{i,10^{th}}$ is simply defined as,
\begin{equation}
    \Sigma_{i,10^{th}} = \frac{10}{\theta_{i,10^{th}}^{2}},
\end{equation}
where $\theta_{i,10^{th}}$ means the angular separation between the object $i$ and its 10$^{th}$-nearest neighbour within the individual redshift bin. 

For sources near the edge of the survey area, the 10$^{th}$-nearest distance cannot provide the true local density in reality, because a part of the neighbouring sources are outside of the survey coverage. Therefore, we performed a correction for the sources which are closer to the edge than to their $10^{th}$-nearest neighbours. We assume the local densities inside and outside the survey are the same. We drew a circle with its centre at the edge galaxy using the 10$^{th}$-nearest distance, and measured the area fraction ($x$) outside the survey coverage. We suppose the $n^{th}$-nearest distance we measured actually is the $m^{th}$-nearest distance in reality (i.e. there are $m$ galaxies in the entire circle, while $n$ galaxies are located in the survey coverage). Due to the assumption of the uniform density, the number of galaxies is proportional to the area fraction ($x$): 
\begin{equation}
    n:1-x = m:1
\end{equation}

Therefore, to obtain the true 10$^{th}$-nearest distance (i.e. $m$=10) of the source near the edge, we use the $n^{th}$-nearest distance instead, where $n=10\times(1-x)$. The $n$ is rounded off to the nearest integer. Note that for some edge sources, their density do not change after the edge correction as $n$ is rounded off to 10. Among the 310577 selected sources, 4732 sources are near the edge, and the corrected local density are adopted for 3690 sources. More details about the local density calculation, and the edge correction are described in Santos et al. 2021 submitted\footnote{They adopted the cosmology from \textit{Wilkinson Microwave Anisotropy Probe} \citep[\textit{WMAP};][]{Komatsu2011} for their density calculation, which has a slight difference from this work.}.

We normalised the local density (denoted as $\Sigma^{*}$ for the normalised local density) with the median density in the individual redshift bin. There exist 7137 sources (about 2 per cent in the selected sample) with $\Sigma^{*}>3$, and 28498 sources (about 9 per cent in the selected sample) with $\Sigma^{*}>2$. We defined those 28498 objects with $\Sigma^{*}>2$ as over-densities. These over-densities were used to search cluster candidates based on the friends-of-friends algorithm in the next section. 

\subsubsection{Friends-of-friends algorithm}
\label{fof}
We used the coordinates, defined as the positions of the HSC detections, and the estimated photo-$z$ to select cluster candidates. 

We selected cluster candidates by applying the friends-of-friends algorithm to all the over-densities. For each object of the over-densities, its friend is defined if (1) the physical projected distance between them is smaller than a given linking length, and (2) the redshift difference (we call it linking redshift, $\Delta z$, hereafter) between them is smaller than 0.032, which is a half of our photo-$z$ NMAD. This value was chosen empirically. Using a large value like 0.065 makes the algorithm link galaxies in wide redshift space, and the redshift of the member galaxies in a generated cluster can deviate largely (e.g. $z_{max}-z_{min}>0.5$), where $z_{max}$ and $z_{min}$ represent the maximum and the minimum redshift of member galaxies. We reasonably believe that such kind of results are highly contaminated. In consequence, we adopted a smaller value (a half of the photo-$z$ NMAD) as the constraint of the linking redshift in the linking process. Note that we did not make this $\Delta z$ value redshift-dependent in order to avoid the widely linking problem described above. We discuss the details about the redshift-dependent linking redshift in Appendix~\ref{dz}. Using the value $\Delta z=0.032$ makes redshift deviations of cluster galaxies more reasonable. For example, in the worst case of our cluster candidates $z_{max}-z_{min}$ is about 0.3, which is comparable to the redshift uncertainty $\pm0.065\times(1+z_{p})$ at $z_{p}=1$.

\begin{figure*}
    \centering
    \includegraphics[width=\columnwidth]{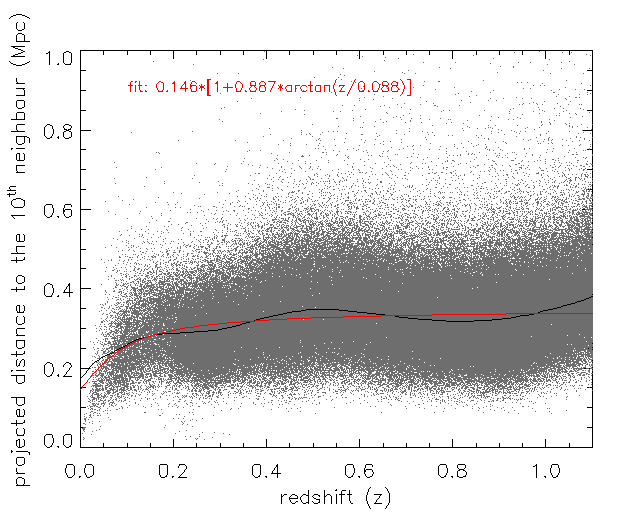}
    \includegraphics[width=\columnwidth]{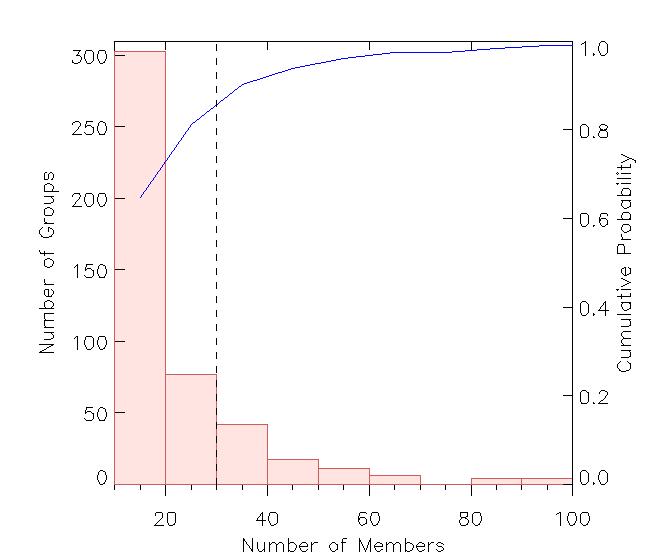}
    \caption{(left panel) The scatter plot of projected distance to the 10$^{th}$-nearest neighbour as a function of redshift for every object. Every grey dot represents a galaxy. The black line is the median value of the distance to the 10$^{th}$-nearest neighbour. The red line is the best-fitting arctangent function to the median values. (right panel) The histogram of group member numbers. The blue curve shows the cumulative probability. The black line indicates where the member number equals 30, the definition of cluster candidates in this work.}
    \label{fig:ll}
\end{figure*}

We used the arctangent function, 
\begin{equation}
f(z) = d_{0} \times [1+\alpha_{0} \arctan{(z/z_{*})}], 
\end{equation}
to estimate the linking length, as suggested by \cite{Tempel2014}. Three parameters $d_{0}$ (in Mpc), $\alpha_{0}$, and $z_{*}$, were determined by a $\chi^{2}$-fitting to the median values of the distances to the 10$^{th}$-nearest neighbours, as shown in Fig.~\ref{fig:ll}. The physical projected distance to the 10$^{th}$-nearest neighbour of every galaxy is plotted in grey. The median values of the distance to the 10$^{th}$-nearest neighbour were calculated for 111 redshift bins ($\pm0.065\times(1+z)$, redshift ranges from 0 to 1.1 with a step of 0.01). The standard error were applied in the $\chi^{2}$-fitting. The best-fitting values of the parameters ($d_{0}$, $\alpha_{0}$, $z_{*}$) are (0.146 Mpc, 0.867, 0.088). 

We start by looking for friends
around the galaxy over-densities using the appropriate linking parameters at redshift of the over-density. Over-densities with at least 10 friends form an initial group. The friends make other new friends in the same way with their own linking length until no new friends are found. The process ends when all over-densities with at least 10 friends are in a group. The algorithm generated 468 groups in total. The right panel of Fig.~\ref{fig:ll} presents numbers of groups in different numbers of galaxy members. There are 165, 88, and 46 groups having more than or equal to 20, 30, and 40 members, corresponding to the cumulative relative frequency of 64.7, 81.2, and 90.2 per cent, respectively. As a compromise between the number of groups and the number of members, the groups with least 30 members are defined to be the galaxy cluster candidates. 

\section{Results}
\label{results}
We found 88 cluster candidates in the \textit{AKARI} NEP field (5.4 deg$^2$) from redshift 0 to 1.1, which are plotted as crosses in Fig.~\ref{fig:sky} and listed in Table~\ref{tab:cluster} (to be discussed in the following subsections). There are a total of 4390 cluster galaxies. We investigate the colours (Section~\ref{colour}) and richness (Section~\ref{richness}) of the cluster candidates. The catalogue and optical colour images of the cluster candidates are provided in Section~\ref{catalogue} and Section~\ref{rgb}, respectively. The reliability of our method and the cluster candidates is examined by false detection (Section~\ref{FD}) and redshift uncertainty (Section~\ref{cp}) tests.

\subsection{Colour studies of clusters}
\label{colour}
\begin{figure*}
	\includegraphics[width=0.65\columnwidth]{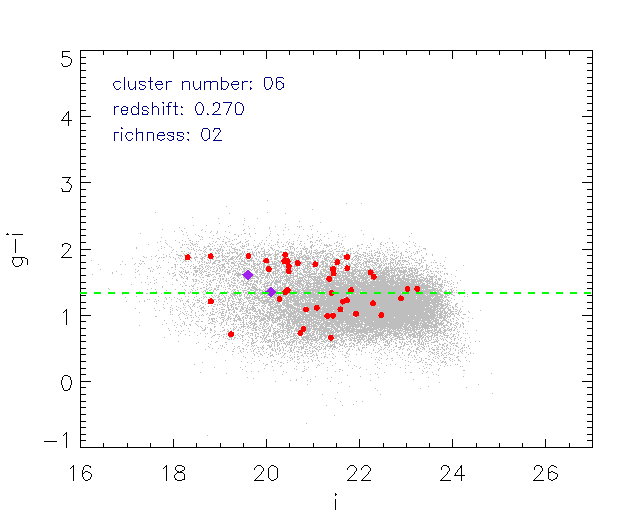}
	\includegraphics[width=0.65\columnwidth]{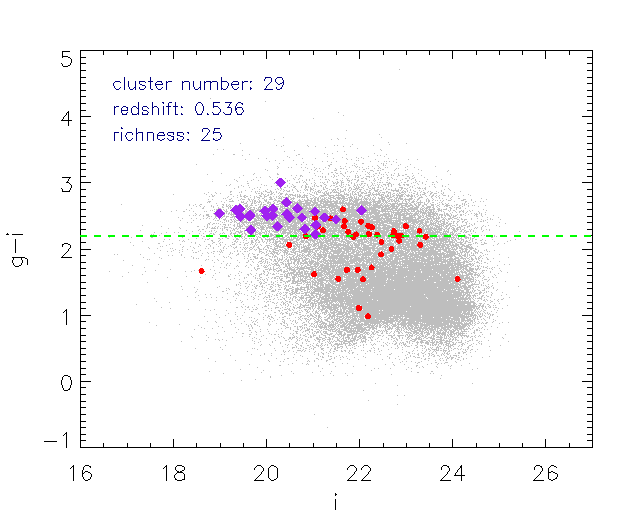}
	\includegraphics[width=0.65\columnwidth]{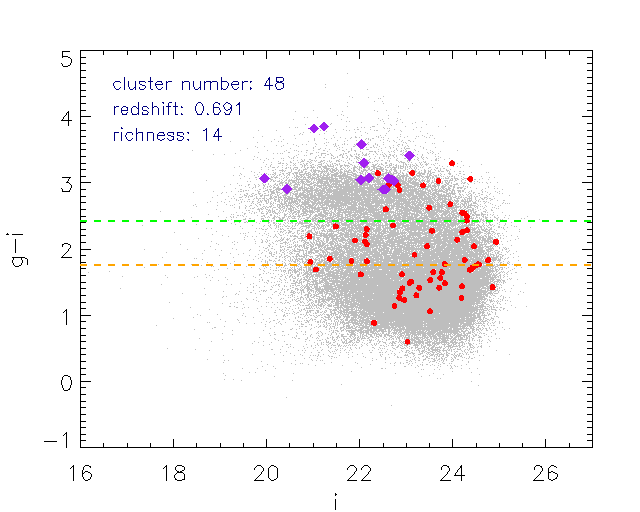}
    \caption{The colour-magnitude diagrams ($g-i$ versus $i$) of the X-ray-matched clusters cl6, cl29, and cl48. The red circles are the cluster galaxies, while the grey dots are the field galaxies in the individual redshift bin. The cluster galaxies that contribute to the cluster richness are plotted as the purple diamonds. The green dashed line shows the colour cut for red galaxies. For cluster candidates at $0.3<z<0.5$ and $0.6<z<0.8$, the red galaxy cut defined by \citet{Lai2016} is indicated by the orange dashed line. }
    \label{fig:CM}
\end{figure*}

We plotted the colour-magnitude diagram ($g-i$ versus $i$) for every cluster candidate identified in this work. The diagrams of all the candidates are available in the online supplementary material. Fig.~\ref{fig:CM} shows the colour-magnitude plots of the X-ray-matched clusters (to be discussed in Section~\ref{Xray}). The cluster galaxies are plotted in red circles, while grey dots represent the non-cluster galaxies (field galaxies hereafter). Since the field galaxies are simply defined as the galaxies not in cluster candidates, this definition includes the galaxies in the groups with the member number less than 30. Nevertheless, the number of these group galaxies (5955) is relatively small compared with the number of all field galaxies (306187), so we believe the statistical contribution from these group galaxies is negligible. The green line shows the colour cut used to define red galaxies, which is an estimate for the cluster richness in the next subsection \ref{richness}.

We empirically define that a galaxy is red if its colour is larger than the colour of the $(60+30z_{cl})$-th percentile in the individual redshift bin $z_{cl}\pm0.065\times(1+z_{cl})$, where $z_{cl}$ is the redshift of the cluster candidate (to be discussed in Section~\ref{catalogue}). Compared with colour cuts for red galaxies from other work, our colour cut is much stricter. For example, \cite{Lai2016} separates red and blue galaxies using $g-i=1.75$ at $0.6<z<0.8$, while our cuts are redder by 0.6 to 0.8 mag in this redshift range. Using their colour cut in our colour-magnitude diagrams (e.g. the right panel of Fig.~\ref{fig:CM}) obviously selects many blue galaxies which are not residing in the galaxy red sequence. We thus believe our redshift-dependent colour cut for red galaxy separation is reliable.\\

\subsection{Richness}
\label{richness}
\begin{figure}
	\includegraphics[width=\columnwidth]{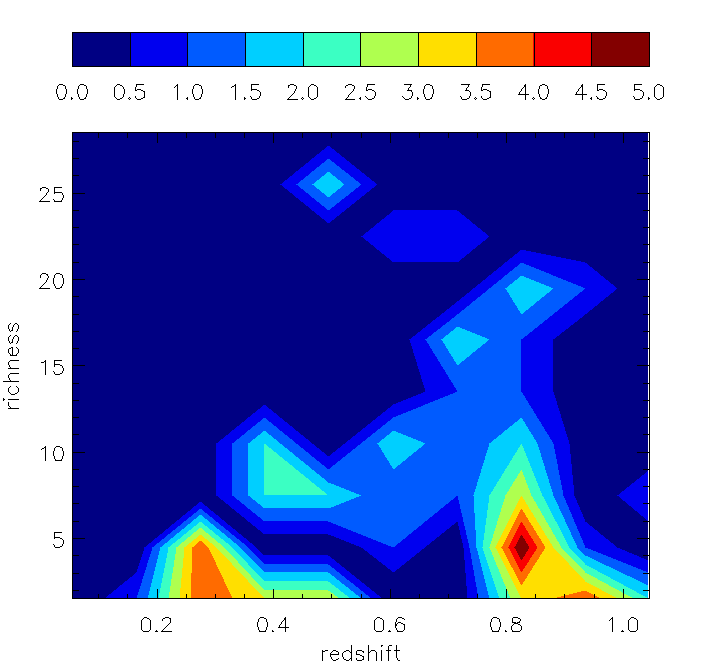}
    \caption{The contour plot presenting the distribution of the cluster candidates in redshift and richness. The contour level is the number density of the cluster candidates.}
    \label{fig:richness}
\end{figure}

The cluster richness provides a way to estimate the cluster masses for clusters detected by optical data. We define the richness of our cluster candidates according to the definition from the studies of HSC CAMIRA galaxy clusters \citep[e.g.][]{Murata2019}, that is, the number of red galaxies with stellar mass larger than $10^{10.2} M_{\odot}$ within 1 $h^{-1}$Mpc from the cluster centre. Moreover, the mass-richness relation has been built by the calibration from weak-lensing magnification \citep{Chiu2020}. In this work, the red galaxies are defined by our own definition in Section \ref{colour}, while the stellar mass is calculated based on the best-fitting stellar models from the SED fitting for the photo-$z$ calculation (Section \ref{photoz}). Fig.~\ref{fig:richness} shows the density map of cluster candidates in redshift and richness as a contour plot. The plot is generated by 10 bins in both redshift range 0 to 1.1 and richness range 0 to 30 with 10 contour levels.

\subsection{Cluster catalogue}
\label{catalogue}
We list the cluster candidates in the \textit{AKARI} NEP field in Table~\ref{tab:cluster}, including the cluster identifier (ID), cluster number (cl), equatorial coordinates (R.A. and Dec.), cluster redshift ($z_{cl}$), maximum normalised local density corresponding to each cluster ($\Sigma^{*}_{max}$), number of member galaxies (N), colour cut for red galaxies ($g-i$ cut), number of red galaxies (N$_{red}$), red galaxy fraction ($f_{red}$), and cluster richness (N$_{richness}$). The cluster number (cl) is the number generated by our friends-of-friends algorithm for identifying groups. We are going to use this number to discuss cluster candidates throughout this paper for convenience. For instance, the first two cluster candidates in Table~\ref{tab:cluster} are denoted as cl18 and cl9. The coordinates and the redshift of the cluster candidates are defined by the member galaxy with the highest normalised local density. If a cluster has member galaxies detected by a spectroscopic observation, the cluster redshift is defined by spec-$z$. In some cases (cl1, cl2, cl14, and cl39), the cluster candidates have two or more members with different spec-$z$. We refer to the spec-$z$ from the brighter/brightest (in $i$ band) one. Two cluster candidates cl13 and cl18 have 2 and 1 members located near the survey edge, respectively. However, all of their densities remain the same after the edge correction because their edge-corrected $n$ are rounded off to 10, as described in Section~\ref{lsd}. The edge correction does not have a direct impact on the density of our cluster candidates, but overall it affects the normalised density and the galaxy over-density in the cluster finding process of this work.

\onecolumn
\begin{longtable}{ccccccccccc}
\label{tab:cluster}\\
\caption{The catalogue of the cluster candidates. The columns are ID, cluster number, right ascension, declination, redshift, maximum normalised local density of the member, number of member galaxies, colour cut for red galaxies, number of red galaxies, red galaxy fraction, and the cluster richness.}\\
 
 \hline
ID & cl & R.A. & Dec. & $z_{cl}$ & $\Sigma^{*}_{max}$ & N & $g-i$ cut & N$_{red}$ & $f_{red}$ & N$_{richness}$ \\
 \hline
 \endfirsthead

 \hline
 \multicolumn{2}{c}{Continuation of Table \ref{tab:cluster}}\\
 \hline
ID & cl & R.A. & Dec. & $z_{cl}$ & $\Sigma^{*}_{max}$ & N & $g-i$ cut & N$_{red}$ & $f_{red}$ & N$_{richness}$ \\
 \hline
 \endhead

 \hline
 \endfoot
 \endlastfoot
HSCJ174647+663904$^{E}$ & 18 & 266.695 & 66.6512 & 0.43 & 11.30 & 85 & 1.90 & 37 & 0.44 & 7\\
HSCJ174720+664409 & 9 & 266.832 & 66.7359 & 0.39 & 9.37 & 45 & 1.73 & 9 & 0.20 & 1\\
HSCJ174725+663506 & 51 & 266.854 & 66.5851 & 0.63 & 8.21 & 47 & 2.31 & 28 & 0.60 & 18\\
HSCJ174954+662637 & 50 & 267.474 & 66.4435 & 0.68 & 5.11 & 49 & 2.40 & 22 & 0.45 & 15\\
HSCJ174954+655724 & 68 & 267.475 & 65.9566 & 0.83 & 9.32 & 33 & 2.56 & 12 & 0.36 & 10\\
HSCJ175012+663115 & 47 & 267.550 & 66.5207 & 0.73 & 5.32 & 51 & 2.48 & 34 & 0.67 & 22\\
HSCJ175024+671005 & 40 & 267.600 & 67.1680 & 0.57 & 7.64 & 32 & 2.24 & 21 & 0.66 & 11\\
HSCJ175026+671446 & 19 & 267.607 & 67.2460 & 0.38 & 4.07 & 32 & 1.66 & 24 & 0.75 & 10\\
HSCJ175032+672156 & 30 & 267.632 & 67.3656 & 0.49 & 5.22 & 41 & 2.10 & 28 & 0.68 & 12\\
HSCJ175137+672904$^{E}$ & 13 & 267.903 & 67.4844 & 0.37 & 4.98 & 54 & 1.65 & 18 & 0.33 & 6\\
HSCJ175156+654457 & 76 & 267.984 & 65.7491 & 0.78$^s$ & 4.38 & 52 & 2.54 & 9 & 0.17 & 8\\
HSCJ175159+671443 & 27 & 267.995 & 67.2453 & 0.41 & 5.38 & 49 & 1.83 & 28 & 0.57 & 10\\
HSCJ175202+661805 & 7 & 268.008 & 66.3014 & 0.28$^s$ & 3.28 & 30 & 1.35 & 6 & 0.20 & 0\\
HSCJ175223+670313 & 23 & 268.095 & 67.0536 & 0.45$^s$ & 8.82 & 53 & 2.00 & 27 & 0.51 & 12\\
HSCJ175234+662346 & 8 & 268.140 & 66.3961 & 0.32 & 10.39 & 44 & 1.46 & 13 & 0.30 & 2\\
HSCJ175311+660356 & 46 & 268.297 & 66.0656 & 0.74 & 11.64 & 56 & 2.50 & 20 & 0.36 & 17\\
HSCJ175321+671957 & 66 & 268.336 & 67.3326 & 0.81 & 4.67 & 36 & 2.55 & 8 & 0.22 & 7\\
HSCJ175330+662245 & 5 & 268.374 & 66.3793 & 0.28$^s$ & 8.99 & 68 & 1.34 & 16 & 0.24 & 5\\
HSCJ175341+663327 & 38 & 268.419 & 66.5576 & 0.52$^s$ & 7.13 & 131 & 2.17 & 52 & 0.40 & 26\\
HSCJ175422+672201 & 72 & 268.592 & 67.3670 & 0.85 & 4.59 & 42 & 2.57 & 15 & 0.36 & 13\\
HSCJ175427+661313 & 59 & 268.611 & 66.2202 & 0.80 & 4.24 & 35 & 2.55 & 4 & 0.11 & 3\\
HSCJ175508+662315 & 14 & 268.782 & 66.3875 & 0.39$^s$ & 5.05 & 36 & 1.74 & 9 & 0.25 & 6\\
HSCJ175524+663517$^X$ & 29 & 268.851 & 66.5881 & 0.54$^s$ & 8.12 & 64 & 2.19 & 45 & 0.70 & 25\\
HSCJ175547+670659 & 56 & 268.946 & 67.1164 & 0.73$^s$ & 9.30 & 55 & 2.48 & 23 & 0.42 & 16\\
HSCJ175554+661022 & 2 & 268.974 & 66.1727 & 0.19$^s$ & 4.02 & 33 & 1.21 & 18 & 0.55 & 2\\
HSCJ175601+662621 & 39 & 269.006 & 66.4392 & 0.53$^s$ & 4.54 & 47 & 2.19 & 16 & 0.34 & 8\\
HSCJ175609+665414 & 22 & 269.036 & 66.9039 & 0.45 & 13.34 & 37 & 1.98 & 3 & 0.08 & 1\\
HSCJ175700+662421 & 3 & 269.249 & 66.4058 & 0.26$^s$ & 5.28 & 55 & 1.31 & 26 & 0.47 & 1\\
HSCJ175704+655047 & 79 & 269.265 & 65.8465 & 0.87 & 5.77 & 30 & 2.58 & 4 & 0.13 & 3\\
HSCJ175715+653920 & 12 & 269.312 & 65.6556 & 0.46 & 3.85 & 37 & 2.03 & 4 & 0.11 & 1\\
HSCJ175717+662912$^X$ & 48 & 269.320 & 66.4867 & 0.69$^s$ & 6.61 & 82 & 2.41 & 30 & 0.37 & 14\\
HSCJ175803+651917 & 11 & 269.512 & 65.3213 & 0.33$^s$ & 5.93 & 47 & 1.47 & 15 & 0.32 & 4\\
HSCJ175828+670144 & 86 & 269.615 & 67.0288 & 0.87 & 5.26 & 30 & 2.58 & 10 & 0.33 & 10\\
HSCJ175846+652726 & 24 & 269.691 & 65.4572 & 0.49 & 8.92 & 134 & 2.10 & 36 & 0.27 & 12\\
HSCJ175846+660853 & 67 & 269.693 & 66.1481 & 0.87 & 6.27 & 41 & 2.58 & 3 & 0.07 & 2\\
HSCJ175856+654053 & 20 & 269.733 & 65.6815 & 0.39$^s$ & 3.78 & 32 & 1.74 & 6 & 0.19 & 3\\
HSCJ175909+661703 & 81 & 269.786 & 66.2843 & 1.00 & 6.90 & 36 & 2.58 & 2 & 0.06 & 2\\
HSCJ175918+651839 & 10 & 269.825 & 65.3108 & 0.28$^s$ & 5.81 & 30 & 1.33 & 13 & 0.43 & 5\\
HSCJ175942+652929 & 41 & 269.926 & 65.4915 & 0.65 & 10.87 & 95 & 2.33 & 37 & 0.39 & 23\\
HSCJ175955+655323 & 44 & 269.979 & 65.8898 & 0.56 & 5.45 & 35 & 2.23 & 24 & 0.69 & 6\\
HSCJ175959+652201 & 16 & 269.996 & 65.3670 & 0.31 & 6.98 & 33 & 1.41 & 7 & 0.21 & 2\\
HSCJ180010+661915 & 80 & 270.041 & 66.3207 & 0.80 & 8.23 & 42 & 2.55 & 4 & 0.10 & 4\\
HSCJ180010+664216 & 78 & 270.041 & 66.7045 & 0.86 & 5.95 & 30 & 2.58 & 5 & 0.17 & 3\\
HSCJ180018+661210 & 49 & 270.076 & 66.2027 & 0.66 & 5.46 & 32 & 2.36 & 14 & 0.44 & 10\\
HSCJ180038+670300 & 84 & 270.158 & 67.0501 & 0.91 & 4.65 & 31 & 2.57 & 4 & 0.13 & 2\\
HSCJ180039+662001 & 28 & 270.164 & 66.3336 & 0.43 & 7.10 & 30 & 1.93 & 25 & 0.83 & 9\\
HSCJ180045+662947 & 69 & 270.187 & 66.4964 & 0.91 & 4.24 & 64 & 2.57 & 8 & 0.12 & 6\\
HSCJ180049+663254 & 65 & 270.206 & 66.5482 & 0.78 & 3.99 & 35 & 2.54 & 6 & 0.17 & 5\\
HSCJ180100+663104 & 45 & 270.249 & 66.5177 & 0.63 & 6.00 & 31 & 2.31 & 8 & 0.26 & 6\\
HSCJ180123+654940 & 15 & 270.347 & 65.8279 & 0.36$^s$ & 3.79 & 32 & 1.60 & 9 & 0.28 & 1\\
HSCJ180124+655214 & 77 & 270.348 & 65.8706 & 0.90 & 4.90 & 31 & 2.57 & 1 & 0.03 & 1\\
HSCJ180144+663714 & 58 & 270.432 & 66.6205 & 0.81 & 11.39 & 64 & 2.55 & 23 & 0.36 & 19\\
HSCJ180147+652627 & 25 & 270.447 & 65.4408 & 0.42 & 5.63 & 44 & 1.89 & 21 & 0.48 & 9\\
HSCJ180157+652856 & 36 & 270.488 & 65.4821 & 0.56 & 5.52 & 56 & 2.22 & 17 & 0.30 & 10\\
HSCJ180201+663400 & 52 & 270.504 & 66.5667 & 0.76 & 8.64 & 88 & 2.53 & 10 & 0.11 & 8\\
HSCJ180210+665859 & 63 & 270.541 & 66.9830 & 0.82 & 6.77 & 32 & 2.56 & 2 & 0.06 & 1\\
HSCJ180221+664354 & 85 & 270.588 & 66.7318 & 0.94 & 4.02 & 40 & 2.57 & 6 & 0.15 & 3\\
HSCJ180228+652929 & 55 & 270.617 & 65.4914 & 0.72 & 4.60 & 31 & 2.47 & 8 & 0.26 & 6\\
HSCJ180242+664858 & 61 & 270.673 & 66.8160 & 0.85 & 8.97 & 84 & 2.57 & 20 & 0.24 & 17\\
HSCJ180248+652606 & 37 & 270.702 & 65.4351 & 0.51 & 6.49 & 38 & 2.13 & 10 & 0.26 & 3\\
HSCJ180256+674845 & 35 & 270.735 & 67.8124 & 0.78 & 7.65 & 66 & 2.54 & 24 & 0.36 & 20\\
HSCJ180300+662646 & 87 & 270.750 & 66.4461 & 0.96 & 5.31 & 31 & 2.58 & 2 & 0.06 & 2\\
HSCJ180311+670830 & 73 & 270.795 & 67.1417 & 1.04 & 4.63 & 51 & 2.62 & 11 & 0.22 & 7\\
HSCJ180318+662428 & 83 & 270.824 & 66.4079 & 0.91 & 5.37 & 31 & 2.57 & 4 & 0.13 & 4\\
HSCJ180323+672822 & 1 & 270.845 & 67.4729 & 0.17$^s$ & 5.06 & 49 & 1.19 & 17 & 0.35 & 3\\
HSCJ180326+655328 & 17 & 270.858 & 65.8910 & 0.33$^s$ & 7.29 & 180 & 1.48 & 106 & 0.59 & 49\\
HSCJ180336+653227 & 34 & 270.898 & 65.5407 & 0.44 & 3.90 & 31 & 1.94 & 18 & 0.58 & 8\\
HSCJ180354+654830 & 21 & 270.976 & 65.8082 & 0.47 & 12.30 & 42 & 2.05 & 16 & 0.38 & 7\\
HSCJ180430+663555 & 82 & 271.126 & 66.5986 & 0.85 & 6.75 & 30 & 2.57 & 3 & 0.10 & 2\\
HSCJ180437+662706 & 42 & 271.154 & 66.4517 & 0.79 & 7.96 & 91 & 2.54 & 24 & 0.26 & 18\\
HSCJ180444+665256 & 32 & 271.184 & 66.8821 & 0.44 & 8.67 & 30 & 1.95 & 2 & 0.07 & 1\\
HSCJ180448+672559 & 33 & 271.201 & 67.4330 & 0.53 & 6.78 & 34 & 2.17 & 6 & 0.18 & 2\\
HSCJ180523+663153 & 53 & 271.345 & 66.5313 & 0.65 & 3.43 & 36 & 2.34 & 6 & 0.17 & 5\\
HSCJ180525+663234 & 62 & 271.354 & 66.5427 & 0.88 & 8.67 & 92 & 2.58 & 17 & 0.18 & 12\\
HSCJ180547+665424 & 4 & 271.444 & 66.9066 & 0.29 & 7.19 & 94 & 1.36 & 20 & 0.21 & 4\\
HSCJ180616+665453 & 31 & 271.568 & 66.9147 & 0.47 & 7.06 & 64 & 2.04 & 27 & 0.42 & 9\\
HSCJ180633+662850 & 70 & 271.636 & 66.4806 & 0.78 & 7.43 & 53 & 2.54 & 11 & 0.21 & 8\\
HSCJ180635+663256 & 88 & 271.644 & 66.5489 & 1.07 & 8.65 & 39 & 2.65 & 3 & 0.08 & 2\\
HSCJ180715+653439$^X$ & 6 & 271.811 & 65.5774 & 0.27$^s$ & 5.21 & 45 & 1.32 & 28 & 0.62 & 2\\
HSCJ180732+664407 & 26 & 271.885 & 66.7353 & 0.45 & 6.34 & 50 & 2.00 & 15 & 0.30 & 9\\
HSCJ180735+664758 & 64 & 271.896 & 66.7995 & 0.88 & 9.79 & 116 & 2.58 & 21 & 0.18 & 19\\
HSCJ180825+665753 & 60 & 272.103 & 66.9646 & 0.77 & 7.53 & 45 & 2.53 & 5 & 0.11 & 4\\
HSCJ180858+663306 & 43 & 272.242 & 66.5516 & 0.62 & 9.10 & 37 & 2.29 & 23 & 0.62 & 7\\
HSCJ180912+660615 & 75 & 272.300 & 66.1041 & 0.89 & 4.86 & 34 & 2.58 & 3 & 0.09 & 2\\
HSCJ181037+665952 & 74 & 272.654 & 66.9979 & 0.87 & 4.95 & 31 & 2.58 & 5 & 0.16 & 4\\
HSCJ181054+665207 & 71 & 272.727 & 66.8685 & 0.85 & 5.47 & 30 & 2.57 & 1 & 0.03 & 0\\
HSCJ181212+660153 & 57 & 273.050 & 66.0313 & 0.78 & 12.72 & 30 & 2.54 & 16 & 0.53 & 12\\
HSCJ181252+661422 & 54 & 273.216 & 66.2394 & 0.87 & 4.91 & 39 & 2.58 & 7 & 0.18 & 5\\
\hline
\end{longtable}
\begin{itemize}
    \item[] $^s$: The cluster has member galaxies detected by spectroscopy, and the redshift is spec-$z$.
    \item[] $^X$: The cluster is matched with a X-ray cluster.
    \item[] $^{E}$: The cluster has member galaxies located near the survey edge.
\end{itemize}

\twocolumn

\subsection{Cluster 3-colour images}
\label{rgb}
\begin{figure*}
	\includegraphics[width=\columnwidth]{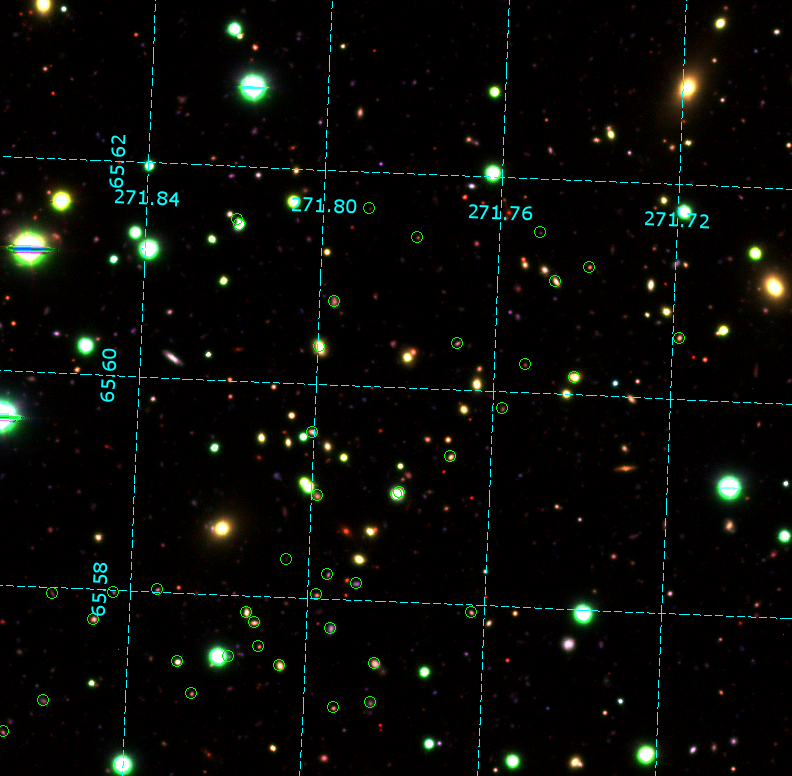}
	\includegraphics[width=\columnwidth]{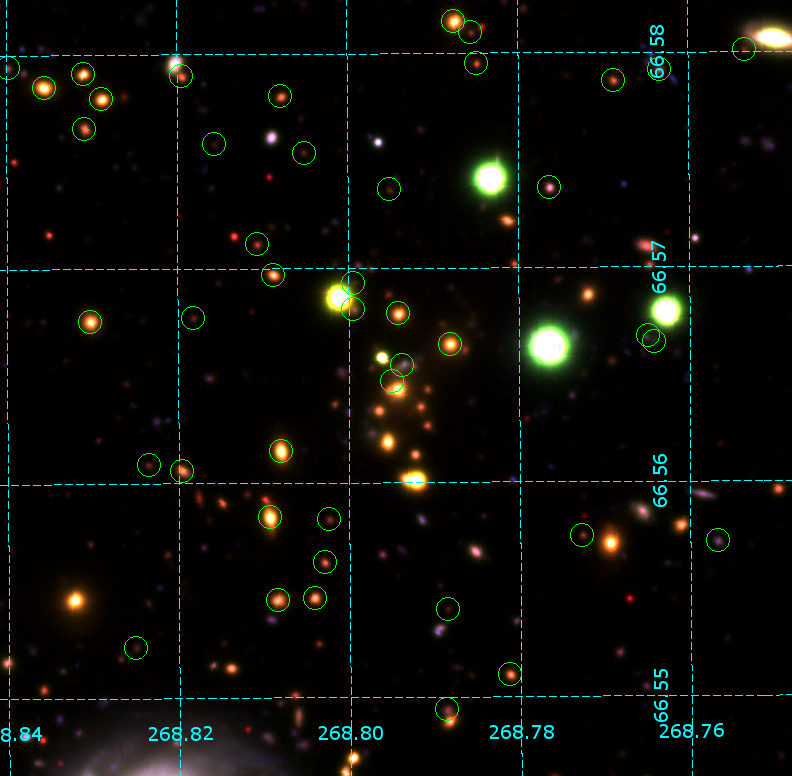}
	\includegraphics[width=\columnwidth]{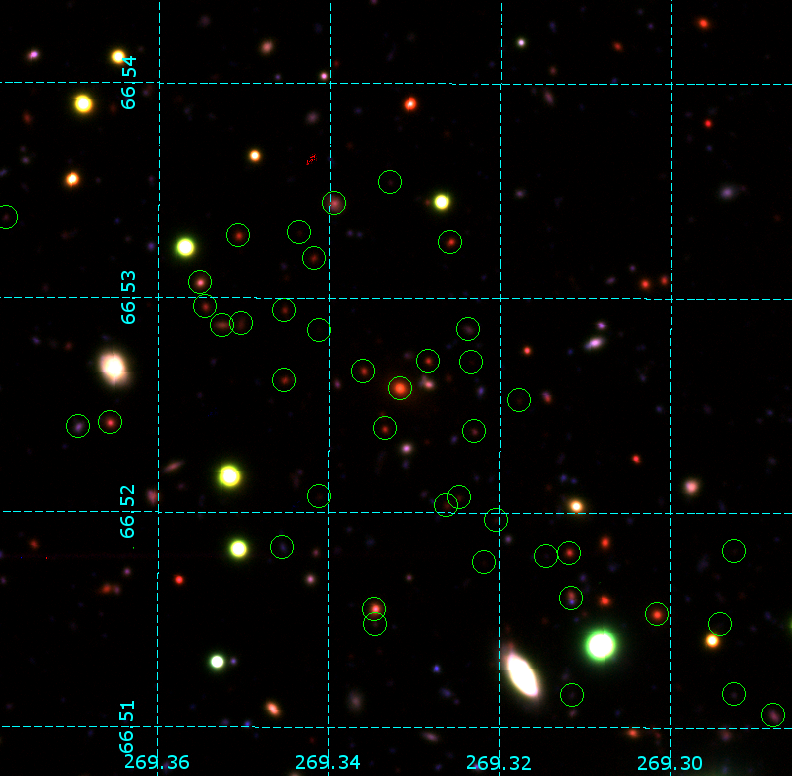}
    \caption{The 3-colour images of the X-ray-matched clusters cl6 (upper left), cl29 (upper right), and cl48 (bottom) created by \textsc{DS9}. HSC $g$-, $r$-, and $i$-band images are used as the blue, green, and red colours in the stacking. The cluster galaxies are circled in green with a radius of 2 arcsec.}
    \label{fig:Xray}
\end{figure*}

We obtained colour composite images of the cluster candidates in the \textit{AKARI} NEP field by mosaicking the HSC $g$-, $r$-, and $i$-band images as blue, green, and red using the \textsc{SAOImageDS9 (DS9)}. We utilised the Subaru HSC coordinate of cluster galaxies to create the input region file of \textsc{DS9} for the cluster galaxies. The colour optical images of the X-ray-matched clusters (discussed in Section~\ref{discussion}) are shown in Fig.~\ref{fig:Xray} as examples. The cluster galaxies are circled in green with the radius of 2 arcsec. The coordinates are shown in cyan dashed lines.   

\subsection{False detection}
\label{FD}
\begin{figure*}
    \includegraphics[width=\columnwidth]{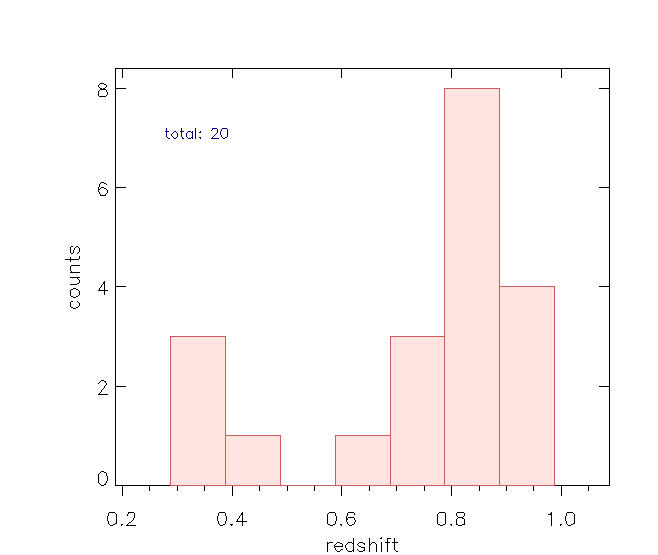}
	\includegraphics[width=\columnwidth]{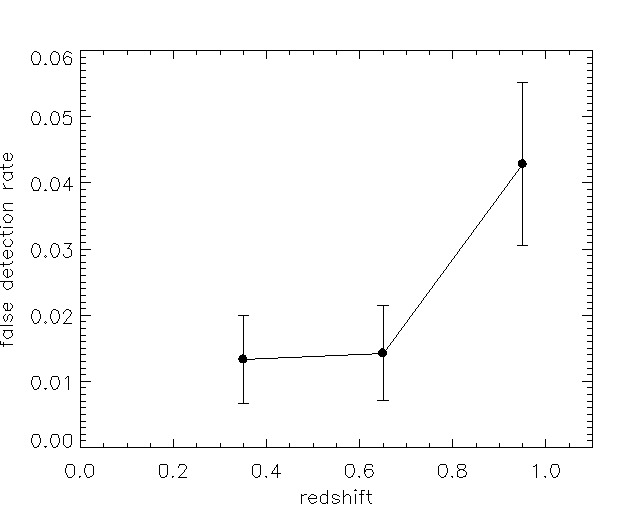}
    \caption{The redshift distribution of the total false detections (left panel, 20 cases from 10 tests) and the false detection rate as a function of redshift (right panel). The errors of the false detection rate are estimated by the Poisson error.}
    \label{fig:fdr}
\end{figure*}

We applied our cluster finding method (i.e. the 10$^{th}$-nearest neighbour and the friends-of-friends algorithm) on mock galaxy catalogues to examine the false detection. The mock galaxies were created through the following steps \citep[e.g.][]{Goto2002,Wen2021}: (1) removing all the cluster galaxies from the selected sample (2) shuffling the redshift values in the remaining sample randomly (keeping the same redshift distribution). We repeated the above steps 10 times and generated 10 different mock galaxy catalogues. The clusters generated in this process are regarded as false detections. The number of false detections in every test is shown in Table~\ref{tab:FD}. We obtained the false detection rate as a function of redshift (Fig.~\ref{fig:fdr}) by dividing the number of the true cluster candidates in each redshift bin (0.2 to 0.5, 0.5 to 0.8, and 0.8 to 1.1). The result indicates that about 1 per cent and 4 per cent false detections happen at $z\leq0.8$ and $z>0.8$, respectively.

We have to remark that this false detection test does not take all kinds of possible false detections into account, but simply testing the sensitivity of the whole cluster finding process to detect chance alignments. Since the groups generated from the friends-of-friends algorithm highly depend on the linking, there may be unrealized false detections. This could be significant merging or fragmentation of groups. Thus, we provide a test on the linking length to check that these two cases do not frequently occur during cluster finding, at least in the projected space.

\begin{figure}
    \includegraphics[width=\columnwidth]{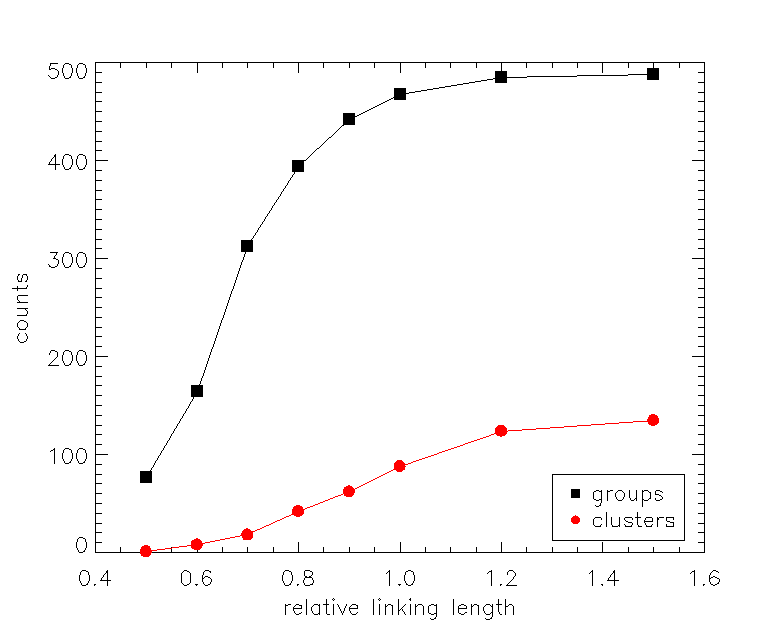}
    \caption{The number of groups (black squares) and clusters (red circles) as functions of the relative linking length, the multiples of original linking length. }
    \label{fig:gp_ll}
\end{figure}

If significant mergers happen, then the number of groups should decrease when a longer linking length is adopted. On the other hand, if significant fragmentations happen due to a short linking length, then numerous groups will be generated. Fig.~\ref{fig:gp_ll} shows that the number of groups/clusters generated by our method increases monotonically with the relative linking length from the multiplier of 0.5 to 1.5. This result implies that the significant mergers/fragmentations do not commonly occur in the linking in the projected space. We interpret that our linking length is properly chosen in the intergalactic scale, so that the significant mergers in the intercluster scale are avoided. Significant fragmentations are not seen either, as the number of groups drops sharply at low relative linking length. This is likely a result of the linking criteria requiring 10 friends, so that the code does not create overflowing groups with only a few members.

\begin{table}
	\centering
	\caption{The number of false detections in every test.}
	\label{tab:FD}
	\begin{tabular}{lr} 
		\hline
		test & false detection\\
		\hline
		1 & 5\\
		2 & 3\\
		3 & 1\\
		4 & 2\\
		5 & 1\\
		6 & 4\\
		7 & 3\\
		8 & 0\\
		9 & 0\\
		10 & 1\\
		\hline
	\end{tabular}
\end{table}

\subsection{Redshift uncertainty test}
\label{cp}
In this section, we test how redshift uncertainty affects our cluster finding by recovering mock clusters under different NMAD conditions. The mock clusters were generated from our detected cluster candidates by changing the redshift values of them and their member galaxies. We assumed all the member galaxies in a mock cluster has a same redshift value but diverges with different redshift errors. In a mock cluster, the redshift value is defined to be the mean redshift ($z_{mean}$) of the detected cluster galaxies. We generated Gaussian-distributed redshift errors ($z_{err}$) according to given redshift NMAD values ($\sigma_{z}$). The redshift of a mock cluster galaxy in a cluster ($z_{mock}$) is written as:

\begin{equation}
\begin{split}
& z_{mock} = z_{mean} + z_{err} \\
& z_{err} = \sigma_{z} \times (1+z_{mean}) \times g(0,1)
\end{split}
\end{equation}
where $g(0,1)$ is a randomly generated number in the Gaussian distribution with the mean of 0 and the standard deviation of 1.

We applied our friends-of-friends algorithm on mock cluster catalogues, and then checked the cluster detection to examine the recovery rate of our method. The recovery rate is defined as the number fraction of the recovered clusters. We further categorised the mock clusters into samples in different redshift ranges (0.2 to 0.5, 0.5 to 0.8, and 0.8 to 1.1) and richness ranges (0 to 5, 6 to 14, and larger than or equal to 15). 10 mock cluster samples were created for each redshift/richness range (i.e. 10 tests were performed for each range). To calculate the recovery rate, the average detection was taken, and the errors were estimated by the standard error. Additionally, two redshift NMAD values (0.03 and 0.06) were used to generate the mock clusters for the redshift uncertainty tests.

\begin{figure*}
	\includegraphics[width=\columnwidth]{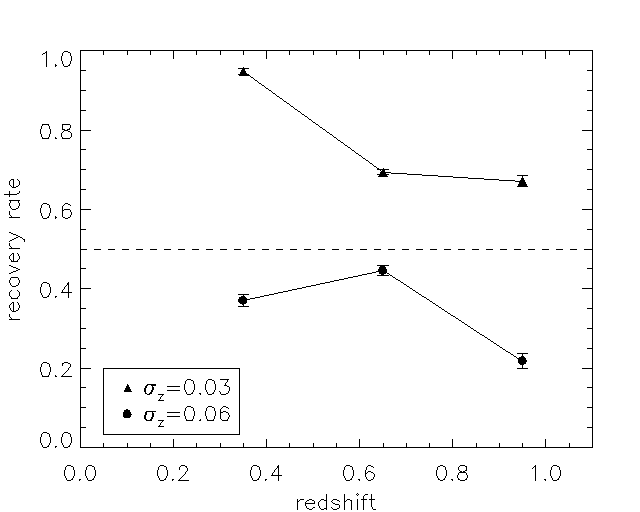}
	\includegraphics[width=\columnwidth]{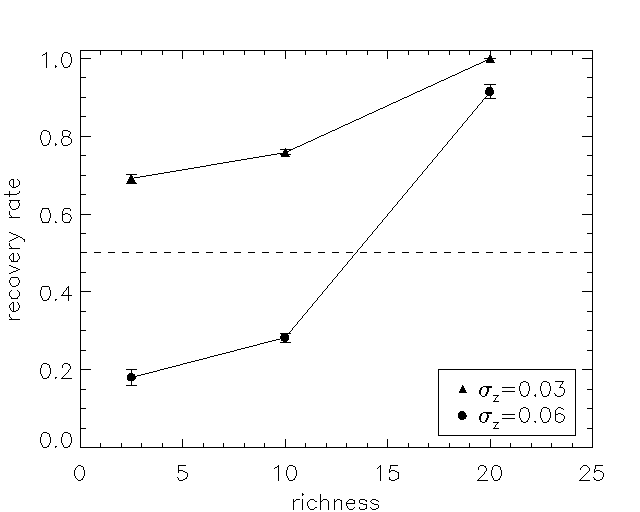}
    \caption{The recovery rate as functions of redshift (left panel) and richness (right panel). The triangles and the circles show the results from the mock cluster sample with different redshift NMAD values, 0.03 and 0.06, respectively. The errors are estimated by the standard error. The dashed horizontal line indicates the 50 per cent recovery rate.}
    \label{fig:cp}
\end{figure*}

The results of the redshift uncertainty tests are shown in the Fig.~\ref{fig:cp}. The triangle and the circle data points represent the recovery rate from different redshift NMAD values. If the redshift NMAD value is 0.06, which is similar to the photo-$z$ performance of our HSC catalogue used in this work, our method achieves the recovery rate of 45 per cent in the intermediate redshift range (0.5 to 0.8). The recovery rate drops at high redshift (>0.8), which can be explained by the high uncertainty (The redshift NMAD 0.06$\times$(1+$z$)>0.1 as $z$>0.8). On the other hand, the recovery rate increases with the richness, which reaches 90 per cent in the high-richness sample (larger than or equal to 15). 

If the redshift NMAD is 0.03, which is a typical value in higher quality samples obtained from fields with more photometry like COSMOS or CFHTLS \citep[e.g.][]{Mobasher2007,Coupun2009}, the 100 per cent recovery rate is achievable in high richness, and even possible to have 65 per cent in the most uncertain high-redshift sample ($z$>0.8). The result of the redshift uncertainty analyses implies that our cluster finding method is feasible, although the current cluster finding is limited by the photo-$z$ accuracy.  

\section{discussion}
\label{discussion}
In this section, we discuss the following topics: a comparison with the known clusters detected by X-ray (Section \ref{Xray}), comparisons between the field and the cluster galaxies (Section \ref{mass}), and cluster finding with the pure photo-$z$ sample (Section \ref{nospecz}). Finally, we discuss some ongoing and future projects which planned to perform observations and contribute to the cluster studies in the \textit{AKARI} NEP field (Section \ref{future}).

\subsection{X-ray clusters}
\label{Xray}
We compared our cluster candidates with the X-ray clusters detected from \textit{ROSAT} and \textit{Chandra}. The X-ray clusters are listed in Table~\ref{tab:Xray}, including the ID, R.A., Dec., redshift, flux, detecting satellite, and match in this work. The \textit{ROSAT} NEP survey \citep{Henry2006} reported 442 X-ray sources, and among them 7 galaxy clusters are located in the \textit{AKARI} NEP field. Two \textit{ROSAT} clusters were recovered by our cluster finding method (matched with cl6 and cl48). We suspect that the other 5 \textit{ROSAT} clusters can not be recovered because they are very nearby ($z$<0.1), or faint in X-ray at low redshift ($\sim5\times10^{-14}$ erg cm$^{-2}$ s$^{-1}$ at $z\sim0.3$). As shown in Fig.~\ref{fig:photoz_dist}, our selected sample contains a very limited number of galaxies at z<0.1, because local galaxies are saturated at bright magnitudes in our HSC deep observations. Therefore, intrinsically it is difficult for us to make groups from nearby galaxies and identify nearby clusters. The X-ray luminosity of a cluster is related to its halo mass \citep[e.g.][]{Stanek2006}, so faint clusters may consist of fewer galaxies or have lower local density, which makes our cluster finding ineffective. On the other hand, 2 clusters were found in the \textit{Chandra} survey in the \textit{AKARI} NEP field. These two clusters are spatially-extended and identified visually in the colour image stacked by 3 X-ray bands \citep{Krumpe2015}. One of them (cl48) is also detected by \textit{ROSAT} as RXJ1757.3+6631, while the other one (cl29) is located around ($\alpha$=17:55:12,$\delta$=+66:33:51). We simply estimated the flux of the latter unknown \textit{Chandra} cluster in Appendix~\ref{flux_est}. Both of the two \textit{Chandra} clusters are recovered through our cluster finding method. The determination of a match is based on the coordinates and the redshift of the X-ray clusters and our detected cluster candidates. The match is defined if a cluster candidate is close to an X-ray cluster within a 2-arcmin radius and a redshift difference of $0.065\times(1+z_{x})$, where $z_{x}$ is the redshift of the X-ray cluster. However, the cluster only detected by \textit{Chandra} has no information in redshift, so it is based only on the coordinates (matched with cl29). In the case of cl29, 2 arcmin corresponds to 0.7 Mpc at redshift 0.54, so the matching radius is probably small enough to claim that the \textit{Chandra} source is correlated to the cluster candidates we found.

\begin{table*}
	\centering
	\caption{The catalogue of the X-ray clusters detected by \textit{ROSAT} and \textit{Chandra} in the \textit{AKARI} NEP field. The columns are ID, right ascension, declination, redshift, flux (0.5 to 2.0 keV), satellite that detected the X-ray cluster, and match with the cluster candidates in this work.}
	\label{tab:Xray}
	\begin{tabular}{lcccccr} 
		\hline
		ID & R.A. & Dec. & redshift & flux (erg cm$^{-2}$ s$^{-1}$) & detecting satellite & match in this work\\
		\hline
		RXJ1806.8+6537 & 271.715 & +65.6294 & 0.26 & 2.74 $\times 10^{-13}$ & \textit{ROSAT} & matched with cl6\\
		RXJ1757.3+6631 & 269.3325 & +65.5275 & 0.69 & 3.40 $\times 10^{-14}$ & both & matched with cl48\\
		RXJ1804.2+6729 & 271.065 & +67.4892 & 0.06 & 4.99 $\times 10^{-14}$ & \textit{ROSAT} & not matched\\
		RXJ1754.7+6623 & 268.6904 & +66.3981 & 0.09 & 1.82 $\times 10^{-13}$ & \textit{ROSAT} & not matched\\
		RXJ1751.5+6719 & 267.8788 & +67.3222 & 0.09 & 1.85 $\times 10^{-13}$ & \textit{ROSAT} & not matched\\
		RXJ1758.9+6520 & 269.74 & +65.3494 & 0.37 & 3.89 $\times 10^{-14}$ & \textit{ROSAT} & not matched\\
		RXJ1808.7+6557 & 272.1817 & +65.9514 & 0.25 & 6.01 $\times 10^{-14}$ & \textit{ROSAT} & not matched\\
		J175511+663354 & 268.797 & +66.565 & 0.54$^{*}$ & 2.94 $\times 10^{-14*}$ & \textit{Chandra} & matched with cl29\\
		\hline
	\end{tabular}
\\$^{*}$: These values are determined in this work.
\end{table*}

\subsection{COSMOS clusters}
\label{cosmos}
To further ensure the robustness and reliability of our method, we provide a rough examination by applying our cluster finding on the COSMOS photo-$z$ catalogue \citep{Ilbert2009}, which has the photo-$z$ uncertainty $\sigma_{z_p}=0.053$ at $0.2<z<1.5$, and compared the result with the COSMOS cluster catalogue \citep{Bellagamba2011}. 

\begin{figure}
	\includegraphics[width=\columnwidth]{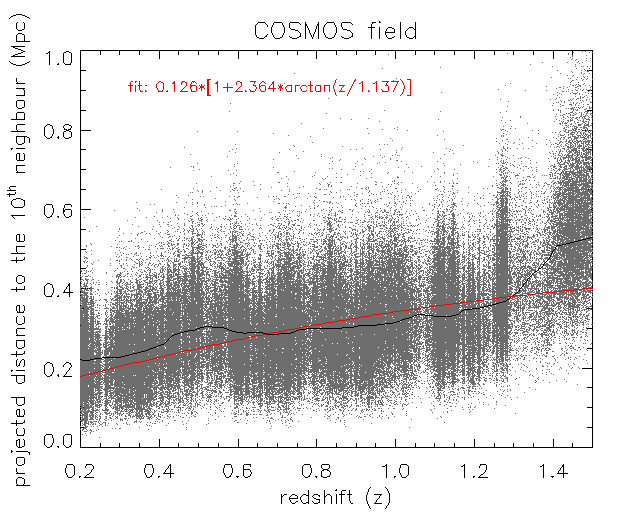}
    \caption{The projected distance to the 10$^{th}$-nearest neighbour as a function of redshift for the photo-$z$ sources in the COSMOS field. The symbols are the same as the left panel of Fig.~\ref{fig:ll}. }
    \label{fig:ll_cosmos}
\end{figure}

We mostly adopted the same approach as this work, that is, calculating the local density and then grouping by the friends-of-friends algorithm. However, since the photo-$z$ uncertainty and the number density of the COSMOS field is different from the ones of the \textit{AKARI} NEP field, we modified those parameters as the same manner in this work. The individual redshift bin, the linking length, and the linking redshift are determined to be $\pm0.05\times(1+z)$, $0.126\times[1+2.364\arctan{(z/1.137)}]$ Mpc, and 0.025, respectively. The linking length was obtained by fitting the median projected distance to the 10$^{th}$-nearest neighbour in Fig.~\ref{fig:ll_cosmos}.

\begin{figure*}
	\includegraphics[width=\columnwidth]{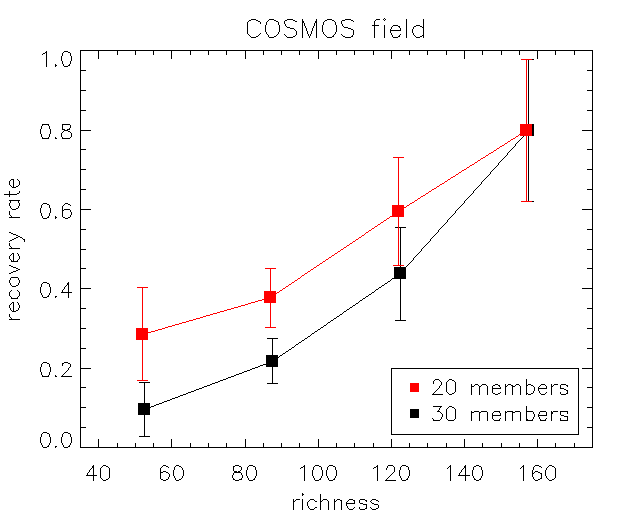}
	\includegraphics[width=\columnwidth]{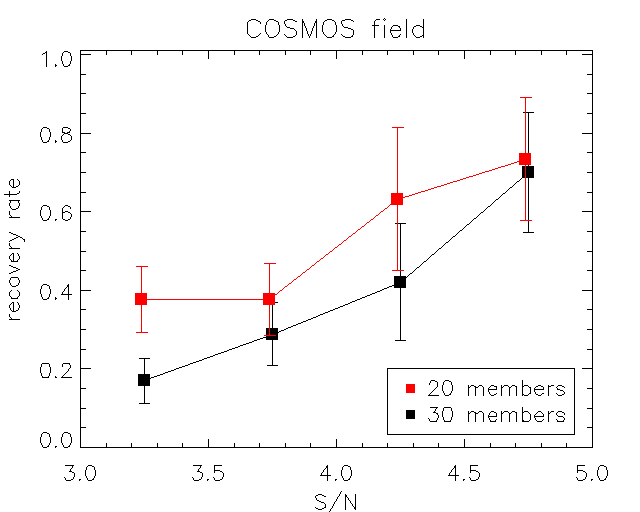}
    \caption{ The recovery rate as functions of richness (left panel) and signal-to-noise ratio (right panel) for the COSMOS data. The red and the black squares show the results from the cluster candidates defined by 20 and 30 members, respectively. The errors are estimated by the Poisson error. }
    \label{fig:cp_cosmos}
\end{figure*}

\begin{figure}
	\includegraphics[width=\columnwidth]{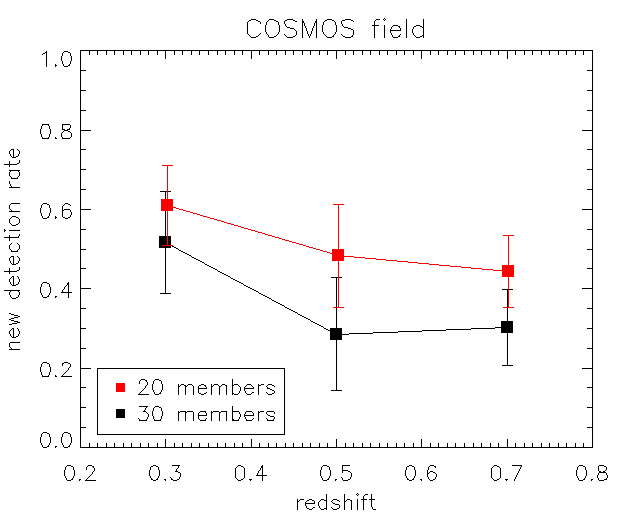}
    \caption{The new detection rate as a function of redshift for the COSMOS data. The red and the black squares show the results from the cluster candidates defined by 20 and 30 members, respectively. The errors are estimated by the Poisson error. }
    \label{fig:ndr}
\end{figure}

Our method selected 139 cluster candidates, and 51 of 147 COSMOS clusters are recovered. The match is defined in the same way as the last section we used for X-ray clusters. We additionally used a more relaxed cluster definition of 20 members for comparison in this section. There are 100 additional relaxed cluster candidates, and 20 of them are matched with the COSMOS clusters. The recovery rate as functions of richness and signal-to-noise ratio (S/N), which are provided by the COSMOS cluster catalogue \citep{Bellagamba2011}, are shown in Fig.~\ref{fig:cp_cosmos}. We also estimated the `new detection rate' for the cluster candidates detected by our method but do not match with COSMOS clusters in Fig.~\ref{fig:ndr}. The recovery rate increases with both richness and S/N, which is consistent with the result of the redshift uncertainty test in Section~\ref{cp} to some extent. Moreover, the low recovery rate at low richness (or low S/N) is relevant to the threshold of member number we define cluster candidates. Nonetheless, loosening the threshold might be risky as the `new detection' increases likewise. These results may indicate that our method is able to capture high-richness clusters, but inherently has a difficulty to detect low-richness clusters.

\subsection{Stellar mass and star formation rate}
\label{mass}
\begin{figure*}
	\includegraphics[width=0.65\columnwidth]{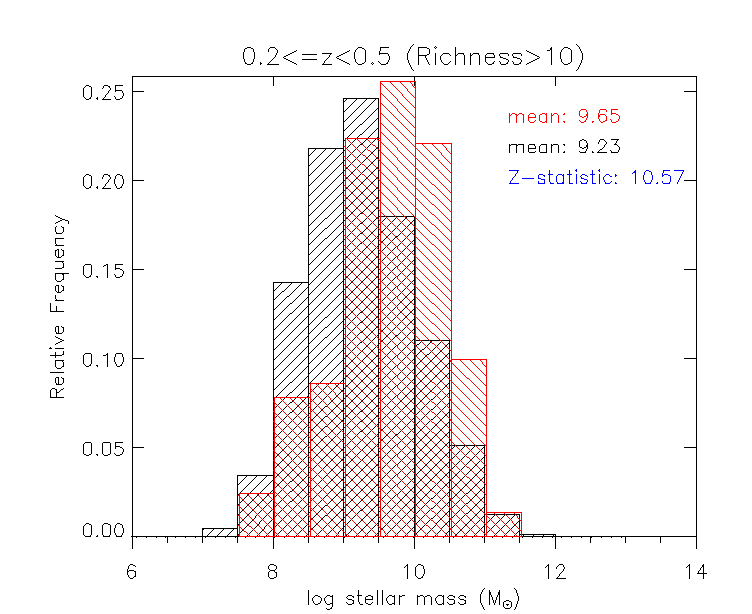}
	\includegraphics[width=0.65\columnwidth]{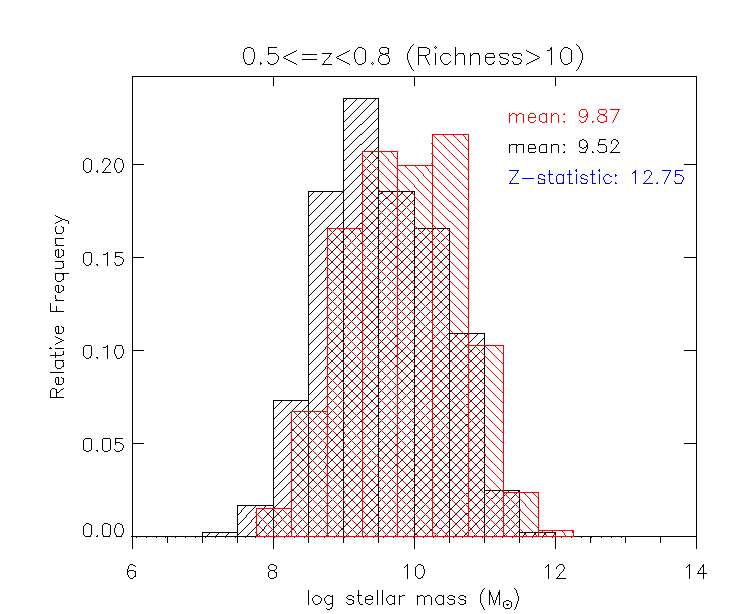}
	\includegraphics[width=0.65\columnwidth]{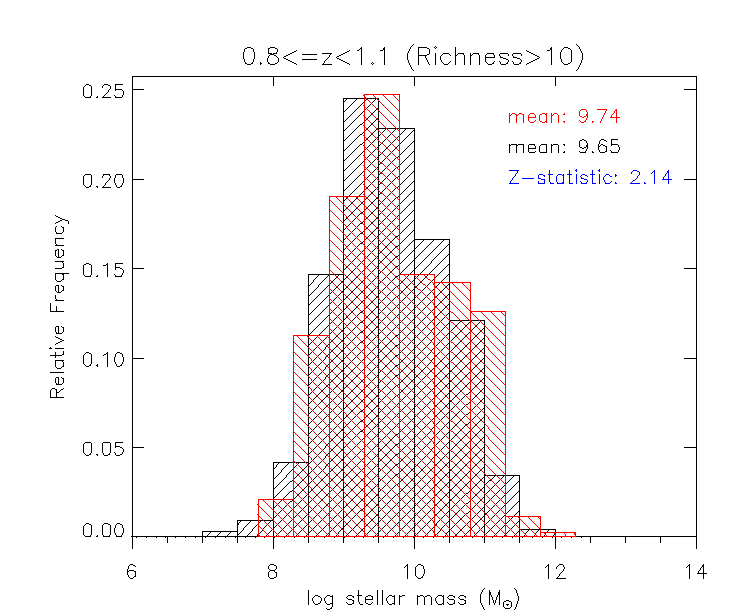}
	\includegraphics[width=0.65\columnwidth]{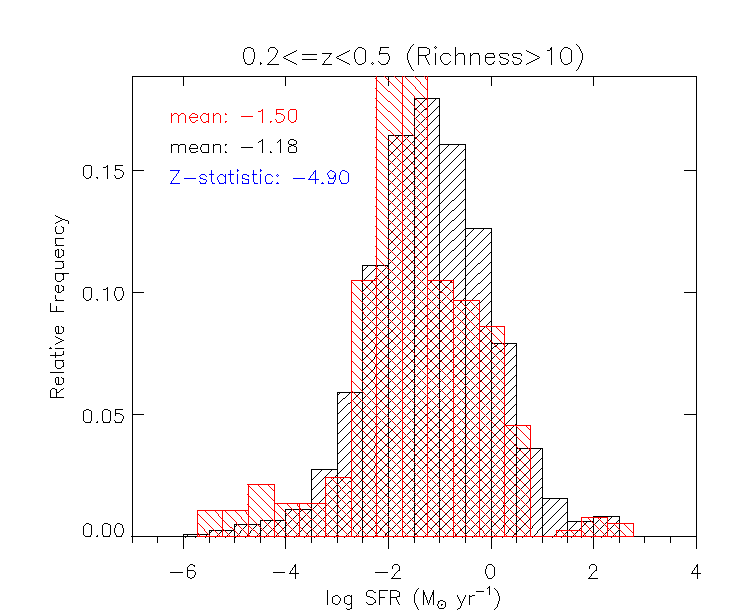}
	\includegraphics[width=0.65\columnwidth]{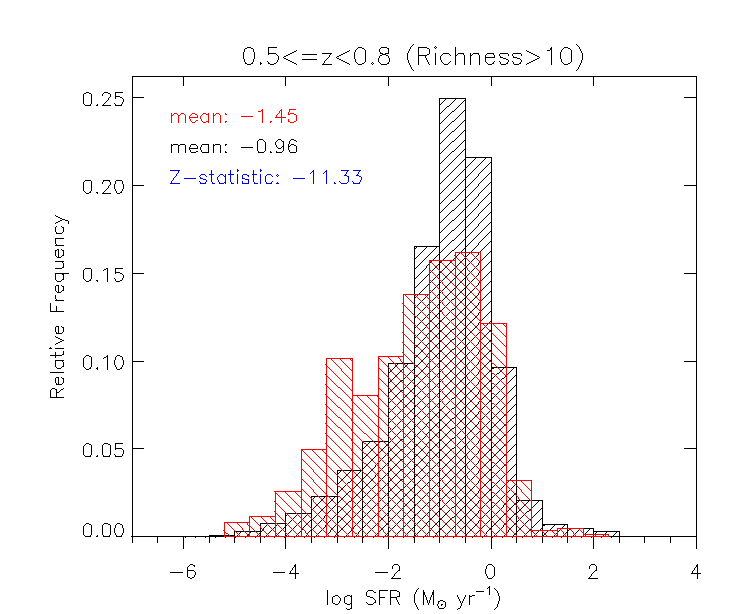}
	\includegraphics[width=0.65\columnwidth]{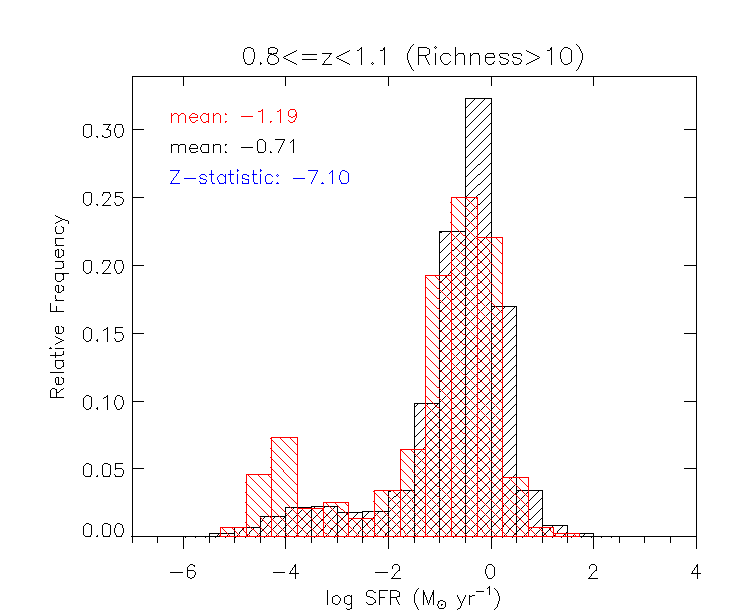}
    \caption{The normalised number distribution of stellar mass and SFR in log scale. The field galaxies and the cluster galaxies are plotted in black and red, respectively. The mean value of each distribution is shown correspondingly.}
    \label{fig:mass}
\end{figure*}

In this section, we compare stellar mass and SFR between the cluster galaxies and the field galaxies. Just like the stellar mass described in Section~\ref{richness}, SFR for every galaxy was also estimated by the SED fitting during the photo-$z$ calculation. Although the SFR is estimated without mid-IR and far-IR photometry, it is still useful for a qualitative comparison of the two galaxy populations. Fig.~\ref{fig:mass} shows the histogram of the stellar mass and the SFR distributions in different redshift ranges (0.2 to 0.5, 0.5 to 0.8, and 0.8 to 1.1). The field galaxies are plotted in black, while the cluster galaxies of the cluster candidates with the richness larger than 10 are plotted in red. The mean values of the stellar mass and SFR are also shown in the figures. The two-sample Z-statistic ($Z$) is used to measure the difference between the two distributions from cluster and field galaxy samples.
\begin{equation}
    Z = \frac{\overline{X_{c}}-\overline{X_{f}}}{\sqrt{\sigma_{X_c}^{2}+\sigma_{X_f}^{2}}},
\end{equation}
where $\overline{X_{c}}$ and $\overline{X_{f}}$ are the mean values of the stellar mass or SFR for the cluster and the field galaxy sample, respectively. The $\sigma_{X_c}$ and $\sigma_{X_f}$ represent the standard errors of the stellar mass or SFR for the cluster and the field galaxy sample, respectively. The result{s} show that the cluster galaxies generally have higher stellar mass and lower SFR compared to field galaxies at $0.2<z\leq1.1$ \citep[e.g.][]{Vulcani2010}, while the difference in stellar mass at the highest redshift bin is not very significant. This indicates that relatively massive cluster galaxies are found by our method. Also, this result may imply that comparing physical properties between cluster and field galaxies statistically using our cluster candidates is a feasible work in the future.

\subsection{Cluster finding without spec-z}
\label{nospecz}
In this section, we examine how much the spec-$z$ improves the redshift-based cluster finding by repeating our cluster finding method with the pure photo-$z$ sample. It resulted that 1 original cluster candidate is missed, and 2 false detection are generated using the pure photo-$z$ sample. The missed cluster candidate is detected as a group with 27 members in the pure photo-$z$ cluster finding, while the 2 false detected cluster candidates are detected as groups with 29 and 26 members in the cluster finding with spec-$z$. A further comparisons in detection numbers, missing rates, and false detection rates of the galaxy over-density, groups, and clusters are listed in the Table~\ref{tab:pure_photoz}. The missing rates and the false detection rates are defined as $\frac{N_{s}-N_{int}}{N_{s}}$ and $\frac{N_{p}-N_{int}}{N_{p}}$, respectively. The $N_{s}$, $N_{p}$, and $N_{int}$ denote the detection number from including spec-$z$, the detection number from pure photo-$z$, and their intersection, respectively. It appears that the results obtained from the pure photo-$z$ catalogue miss the true structures by 1.1 to 1.7 per cent, and contain the false detection rates of 1.6 to 3.3 per cent. This may imply that with the aid of a small fraction of spec-$z$ sources (even less than 1 per cent of the whole selected sample), an improvement on the cluster finding can be marginally seen.

\begin{table*}
	\centering
	\caption{The table presents the detection numbers, missing rate, and false detection rate of galaxy over-densities, groups, group galaxies, clusters, and cluster galaxies. The left three columns show the detection numbers obtained from the cluster finding process with spec-$z$ ($N_{s}$), with pure photo-$z$ ($N_{p}$), and their intersection ($N_{int}$). The fourth and the fifth column from left are the missing rate ($\frac{N_{s}-N_{int}}{N_{s}}$) and the false detection rate ($\frac{N_{p}-N_{int}}{N_{p}}$) of the cluster finding using the pure photo-$z$ catalogue }
	\label{tab:pure_photoz}
	\begin{tabular}{lccccr} 
		\hline
		 & spec-$z$ ($N_{s}$) & pure photo-$z$ ($N_{p}$) & intersection ($N_{int}$) & missing rate ($\frac{N_{s}-N_{int}}{N_{s}}$) & false detection rate ($\frac{N_{p}-N_{int}}{N_{p}}$)\\
		\hline
		over-density & 28498 & 28559 & 28093 & 1.4$\%$ & 1.6$\%$\\
		group & 468 & 472 & 463 & 1.1$\%$ & 1.9$\%$\\
		group galaxy & 10345 & 10446 & 10164 & 1.7$\%$ & 2.7$\%$\\
		cluster & 88 & 89 & 87 & 1.1$\%$ & 2.2$\%$\\
		cluster galaxy & 4390 & 4461 & 4314 & 1.7$\%$ & 3.3$\%$\\
		\hline
	\end{tabular}
\end{table*}

\subsection{Future prospects}
\label{future}
In this section, we describe some large projects in the near future which may potentially help the study of galaxy clusters in the \textit{AKARI} NEP field. We envision the cluster candidates provided in this work will be the observation targets in these future projects and hopefully enhance the chance to find real clusters. Also, we will be able to confirm how many cluster and cluster galaxy candidates are true with the aid of these follow-up observations, and then we can conduct further cluster studies or improve the cluster finding.

The NEP is one of the major targets of \textit{Euclid} mission \citep{Laureijs2011}, which also has a plan to carry out a large-area optical and mid-IR survey for probing galaxy clusters \citep{Euclid2019}. The cluster candidates found in this work will be useful as references, and the results of \textit{Euclid} will also help to advance
cluster research in the \textit{AKARI} NEP field.

The X-ray detector \textit{eROSITA}, which is expected to carry out an all-sky survey and detect more than $10^{5}$ galaxy clusters \citep{Merloni2012}, now has now been launched and is in operation. Having more detected clusters in the \textit{AKARI} NEP field not only directly provides the targets for cluster studies, but also assists the development of cluster detection like this work.

As we have shown in Section~\ref{cp} and~\ref{nospecz}, the redshift accuracy is a key variable for our cluster finding. The accurate redshift can only be achieved by the large number of spectroscopic reference objects. Larger spectroscopic surveys on the NEP are planned using Subaru Prime Focus Spectrograph \citep[PFS;][]{Tamura2016}, which is able to perform a spectral observation of 2400 objects simultaneously. Subaru PFS is expected to start the operation in 2023. We expect that using Subaru PFS can dramatically increase the sample size of spectroscopic objects in the \textit{AKARI} NEP field in the near future.

TolTEC \citep{Bryan2018} is a millimeter-wave camera mounted on the focal plane of the 50-meter Large Millimeter Telescope (LMT). It is able to provide sensitive three-band imaging at 2.0, 1.4, and 1.1 mm with the beamsize of 9.5, 6.3, and 5 arcseconds, respectively. TolTEC also has fast mapping speed at 69, 20, 12, deg$^{2}$mJy$^{-2}$hr$^{-1}$ in maximum, so that it can conduct the 100-deg$^{2}$ Large Scale Structure Survey. As one of the main scientific goals of TolTEC, this survey is expected to probe galaxy clusters via the SZ effect using TolTEC's advantageous three bands.

\section{Summary}
\label{summary}
This work develops an approach for searching galaxy cluster candidates using only positional information based on optical detection of galaxies with as few requisites and selection biases as possible. The cluster finding process applies the friends-of-friends algorithm to galaxy over-densities. The over-densities are selected based on the normalised local surface density of every galaxy determined by the 10$^{th}$-nearest neighborhood within the individual redshift bin. The linking process of the algorithm is designed to utilise the redshift-dependent linking length and the constant linking redshift. The galaxies having more than 10 friends are made into groups, and the groups with more than or equal to 30 members are defined to be the cluster candidates. All the parameters used in the cluster finding are listed below.
\begin{itemize}
    \item k$^{th}$-nearest neighbourhood: 10$^{th}$
    \item individual redshift bin: $\pm0.065\times(1+z)$
    \item over-density: $>2\times$ median
    \item linking length: $0.146\times[1+0.887\arctan{(z/0.088)}]$ Mpc
    \item linking redshift: 0.032
    \item grouping criterion: 10 friends
    \item cluster definition: $\geq$ 30 members
\end{itemize}

As a result, a catalogue of 88 galaxy cluster candidates in the \textit{AKARI} NEP field is presented. Bright X-ray clusters at intermediate redshift can be successfully recovered through our method. The reliability analyses suggest that even under the worst case of the photo-$z$ accuracy, the cluster finding in this work still has fair performance in the recovery rate of 40 per cent and the false detection rate of 1 per cent at redshift 0.2 to 0.8. The cluster candidates with richness larger than or equal to 15 especially have a great recovery rate of 90 per cent. Applying our cluster finding method on the COSMOS field results that 80 per cent of high-richness and high-S/N clusters are recovered. Although some cluster candidates provided in this work might be contaminated by field galaxies, our method is still able to detect some high confidence cluster galaxy candidates for further studies. The comparison with the pure photo-$z$ sample suggest that having a small fraction of spec-$z$ sources slightly improves the cluster finding. Additionally, this method is useful as a first screening for finding galaxy clusters from a large-area survey, efficiently providing potential candidates for further X-ray, radio, or any cluster studies. The NEP is crucial area where many ongoing or future projects like \textit{eROSITA} and \textit{Euclid} are aiming at. The method and cluster candidates provided by this work will be good references for them. 

\section*{Acknowledgements}

We thank the anonymous reviewer for reading the manuscript carefully and providing useful suggestions to make this paper complete. T.-C. Huang thanks Fuda Nguyen and Takuya Yamashita for providing inspiring discussions and thoughts at the early stage of this work. This research is based on observations with \textit{AKARI}, a JAXA project with the participation of ESA. This work is based on data collected at Subaru Telescope, which is operated by the National Astronomical Observatory of Japan (NAOJ). This work is based on observations obtained with MegaPrime/MegaCam, a joint project of the Canada-France-Hawaii Telescope (CFHT) and CEA/DAPNIA, at CFHT which is operated by the National Research Council (NRC) of Canada, the Institut National des Sciences de l'Univers of the Centre National de la Recherche Scientifique (CNRS) of France, and the University of Hawaii. The observations at the CFHT were performed with care and respect from the summit of Maunakea which is a significant cultural and historic site. This work used high-performance computing facilities operated by the Centre for Informatics and Computation in Astronomy (CICA) at National Tsing Hua University (NTHU). This equipment was funded by the Ministry of Education of Taiwan, the Ministry of Science and Technology of Taiwan (MOST), and NTHU. T. Goto is supported by the grant 108-2628-M-007-004-MY3 from MOST. T. Hashimoto is supported by CICA at NTHU through a grant from the Ministry of Education of the Republic of China (Taiwan). W. J. Pearson has been supported by the Polish National Science Center project UMO-2020/37/B/ST9/00466. H. Shim acknowledges the support from the National Research Foundation of Korea (NRF) grant, No. 2021R1A2C4002725, funded by Korea government (MSIT). T. Miyaji is supported by UNAM-DGAPA (PASPA and PAPIIT IN 111319) and CONACyT Grant 252531. H. S. Hwang was supported by the New Faculty Startup Fund from Seoul National University.

\section*{Data availability}
The data underlying this article will be shared on reasonable request to the corresponding author.





\begin{thebibliography}{99}
\bibitem[\protect\citeauthoryear{Arnouts et al.}{1999}]{Arnouts1999}
Arnouts S. et al., 1999, MNRAS, 310, 540
\bibitem[\protect\citeauthoryear{Bellagamba et al.}{2011}]{Bellagamba2011}
Bellagamba F., Maturi M., Hamana T., Meneghetti M., Miyazaki S., Moscardini L., 2011, MNRAS, 413, 1145
\bibitem[\protect\citeauthoryear{Bleem et al.}{2015}]{Bleem2015}
Bleem L. E. et al, 2015, ApJ, 216, 27
\bibitem[\protect\citeauthoryear{Bohlin et al.}{1995}]{Bohlin1995}
Bohlin R. C., Colina L., Finley D. S., 1995, AJ, 110, 1316
\bibitem[\protect\citeauthoryear{B\"oehringer et al.}{2001}]{Boehringer2001}
B\"ohringer H. et al., 2001, A$\&$A, 369, 826
\bibitem[\protect\citeauthoryear{Bryan et al.}{2018}]{Bryan2018}
Bryan S. et al., 2018, Proc. SPIE Conf. Ser. Vol. 10708, p. 107080J
\bibitem[\protect\citeauthoryear{Chabrier et al.}{2000}]{Chabrier2000}
Chabrier G., Baraffe I., Allard F., Hauschildt P., 2000, ApJ, 542, 464
\bibitem[\protect\citeauthoryear{Chen et al.}{2021}]{Chen2021}
Chen B. H. et al., 2021, MNRAS, 501, 3951
\bibitem[\protect\citeauthoryear{Chiu et al.}{2020}]{Chiu2020}
Chiu I.-N., Umetsu K., Murata R., Medezinski E., Oguri M., 2020, MNRAS, 495, 428
\bibitem[\protect\citeauthoryear{Coe et al.}{2013}]{Coe2013}
Coe D. et al, 2013, ApJ, 762, 32
\bibitem[\protect\citeauthoryear{Coupon et al.}{2009}]{Coupun2009}
Coupun J. et al, 2009, A$\&$A, 500, 981
\bibitem[\protect\citeauthoryear{D\'iaz Tello et al.}{2017}]{Diaz2017}
D\'iaz Tello J. et al, 2017, A$\&$A, 604, A14
\bibitem[\protect\citeauthoryear{Dressler}{1980}]{Dressler1980}
Dressler A., 1980, ApJ, 236, 351
\bibitem[\protect\citeauthoryear{Ebeling and Wiedenmann}{1993}]{EW1993}
Ebeling H., Wiedenmann G., 1993, Phys. Rev. E, 47, 704
\bibitem[\protect\citeauthoryear{Euclid Collaboration}{2019}]{Euclid2019}
Euclid Collaboration, 2019, A$\&$A, 627, A23
\bibitem[\protect\citeauthoryear{Fruscione et al.}{2006}]{Fruscione2006}
Fruscione A. et al., 2006, Proc. SPIE Conf. Ser. Vol. 6270, p. 62701V
\bibitem[\protect\citeauthoryear{Goto et al.}{2002}]{Goto2002}
Goto T. et al., 2002, AJ, 123, 1807
\bibitem[\protect\citeauthoryear{Goto et al.}{2003}]{Goto2003}
Goto T., Yamauchi C., Fujita Y., Okamura S., Sekiguchi M., Smail I., Bernardi M., Gomez P. L., 2003, MNRAS, 346, 601
\bibitem[\protect\citeauthoryear{Goto et al.}{2008}]{Goto2008}
Goto T. et al., 2008, PASJ, 60, S531 
\bibitem[\protect\citeauthoryear{Goto et al.}{2017}]{Goto2017}
Goto T. et al., 2017, Publ. Korean Astron. Soc., 32, 225
\bibitem[\protect\citeauthoryear{Henry et al.}{2006}]{Henry2006}
Henry J. P. et al., 2006, ApJS, 162, 304
\bibitem[\protect\citeauthoryear{Hilton et al.}{2020}]{Hilton2020}
Hilton M. et al., 2020, preprint (\href{https://arxiv.org/abs/2009.11043}{arXiv:2009.11043})
\bibitem[\protect\citeauthoryear{Ho et al.}{2021}]{Ho2021}
Ho S. C.-C. et al., 2021, MNRAS, 502, 140
\bibitem[\protect\citeauthoryear{Hong et al.}{2012}]{Hong2012}
Hong T., Han J. L., Wen Z. L., Sun L., Zhan H., 2012, ApJ, 749, 81
\bibitem[\protect\citeauthoryear{Hong et al.}{2016}]{Hong2016}
Hong T., Han J. L., Wen Z. L., 2016, ApJ, 826, 154
\bibitem[\protect\citeauthoryear{Huang et al.}{2017}]{Huang2017}
Huang T.-C., Goto T., Hashimoto T., Oi N., Matsuhara H., 2017, MNRAS, 471, 4239
\bibitem[\protect\citeauthoryear{Huang et al.}{2020}]{Huang2020}
Huang T.-C. et al., 2020, MNRAS, 498, 609
\bibitem[\protect\citeauthoryear{Hung et al.}{2020}]{Hung2020}
Hung D. et al., 2020, MNRAS, 491, 5524
\bibitem[\protect\citeauthoryear{Hwang et al.}{2012}]{Hwang2012}
Hwang H. S., Park. C., Choi Y.-Y., 2012, A$\&$A, 538, A15
\bibitem[\protect\citeauthoryear{Huchra $\&$ Geller}{1982}]{HG1982}
Huchra J. P., Geller M. J., 1982, ApJ, 257, 423
\bibitem[\protect\citeauthoryear{Ilbert et al.}{2006}]{Ilbert2006}
Ilbert O. et al., 2006, A$\&$A, 457, 841
\bibitem[\protect\citeauthoryear{Ilbert et al.}{2009}]{Ilbert2009}
Ilbert O. et al., 2009, ApJ, 690, 1236
\bibitem[\protect\citeauthoryear{Kim et al.}{2018}]{Kim2018}
Kim H. K., Malkan M. A., Oi N., Burgarella D., Buat V., Takagi T., Matsuhara H., 2018, in Ootsubo T., Yamamura I., Murata K., Onaka T., eds, The Cosmic Wheel and the Legacy of the AKARI Archive: From Galaxies and Stars to Planets and Life, p. 371
\bibitem[\protect\citeauthoryear{Kim et al.}{2012}]{Kim2012}
Kim S. J. et al., 2012, A$\&$A, 548, A29
\bibitem[\protect\citeauthoryear{Kim et al.}{2019}]{Kim2019}
Kim S. J. et al., 2019, PASJ, 71, 11
\bibitem[\protect\citeauthoryear{Kim et al.}{2021}]{Kim2021}
Kim S. J. et al., 2021, MNRAS, 500, 4078
\bibitem[\protect\citeauthoryear{Ko et al.}{2012}]{Ko2012}
Ko J. et al., 2012, ApJ, 745, 181
\bibitem[\protect\citeauthoryear{Komatsu et al.}{2011}]{Komatsu2011}
Komatsu E. et al., 2011, ApJS, 192, 18
\bibitem[\protect\citeauthoryear{Krumpe et al.}{2015}]{Krumpe2015}
Krumpe M. et al., 2015, MNRAS, 446, 911
\bibitem[\protect\citeauthoryear{Lai et al.}{2016}]{Lai2016}
Lai C.-C. et al., 2016, ApJ, 825, 40
\bibitem[\protect\citeauthoryear{Laureijs et al.}{2011}]{Laureijs2011}
Laureijs R. et al., 2011, Euclid Definition Study Report, arXiv e-prints (\href{https://arxiv.org/abs/1110.3193}{arXiv:1110.3193})
\bibitem[\protect\citeauthoryear{Matsuhara et al.}{2006}]{Matsuhara2006}
Matsuhara H. et al., 2006, PASJ, 58, 673
\bibitem[\protect\citeauthoryear{Merloni et al.}{2012}]{Merloni2012}
Merloni A. et al., 2012, eROSITA Science Book, arXiv e-prints (\href{https://arxiv.org/abs/1209.3114}{arXiv:1209.3114})
\bibitem[\protect\citeauthoryear{Miller et al.}{2003}]{Miller2003}
Miller C. J., Nichol R. C., G\'omez P. L., Hopkins A. M., Bernardi M., 2003, ApJ, 597, 142
\bibitem[\protect\citeauthoryear{Miyazaki et al.}{2012}]{Miyazaki2012}
Miyazaki S. et al., 2012, Proc. SPIE Conf. Ser. Vol. 8446, p. 84460Z
\bibitem[\protect\citeauthoryear{Mobasher et al.}{2007}]{Mobasher2007}
Mobasher B. et al., 2007, ApJS, 172, 117
\bibitem[\protect\citeauthoryear{Murakami et al.}{2007}]{Murakami2007}
Murakami H. et al., 2007, PASJ, 59, S369
\bibitem[\protect\citeauthoryear{Murata et al.}{2013}]{Murata2013}
Murata K. et al., 2013, A$\&$A, 559, A132
\bibitem[\protect\citeauthoryear{Murata et al.}{2014}]{Murata2014}
Murata K. et al., 2014, A$\&$A, 566, A136
\bibitem[\protect\citeauthoryear{Murata et al.}{2019}]{Murata2019}
Murata R. et al., 2019, PASJ, 71, 107
\bibitem[\protect\citeauthoryear{Muzzin et al.}{2013}]{Muzzin2013}
Muzzin A., Wilson G., Demarco R., Lidman C., Nantais J., Hoekstra H., Yee H. K. C., Rettura A., 2013, ApJ, 767, 39
\bibitem[\protect\citeauthoryear{Nayyeri et al.}{2018}]{Nayyeri2018}
Nayyeri H. et al., 2018, ApJ, 234, 38
\bibitem[\protect\citeauthoryear{Oguri et al.}{2018}]{Oguri2018}
Oguri M. et al., 2018, PASJ, 70, S20
\bibitem[\protect\citeauthoryear{Oi et al.}{2014}]{Oi2014}
Oi N. et al., 2014, A$\&$A, 566, A60
\bibitem[\protect\citeauthoryear{Oi et al.}{2018}]{Oi2018}
Oi N., Goto T., Malkan M. A., Pearson C., Matsuhara H., 2018, PASJ, 69, 70
\bibitem[\protect\citeauthoryear{Oi et al.}{2021}]{Oi2021}
Oi N. et al., 2021, MNRAS, 500, 5024
\bibitem[\protect\citeauthoryear{Onaka et al.}{2007}]{Onaka2007}
Onaka T. et al., 2007, PASJ, 59, S401
\bibitem[\protect\citeauthoryear{Ohyama et al.}{2018}]{Ohyama2018}
Ohyama Y. et al., 2018, A$\&$A, 618, A101
\bibitem[\protect\citeauthoryear{Park $\&$ Hwang}{2009}]{Park2009}
Park C., Hwang H. S., 2009, ApJ, 699, 1595
\bibitem[\protect\citeauthoryear{Pickles}{1998}]{Pickles1998}
Pickles A. J., 1998, PASP, 110, 863
\bibitem[\protect\citeauthoryear{Planck Collaboration XXIX}{2014}]{Planck2014}
Planck Collaboration XXIX, 2014, A$\&$A, 571, A29
\bibitem[\protect\citeauthoryear{Planck Collaboration XXVII}{2016}]{Planck2016}
Planck Collaboration XXVII, 2016, A$\&$A, 594, A27
\bibitem[\protect\citeauthoryear{Poliszczuk et al.}{2019}]{Poliszczuk2019}
Poliszczuk A. et al., 2019, PASJ, 71, 65
\bibitem[\protect\citeauthoryear{Poliszczuk et al.}{2021}]{Poliszczuk2021}
Poliszczuk A. et al., 2021, A$\&$A, in press
\bibitem[\protect\citeauthoryear{Rykoff et al.}{2014}]{Rykoff2014}
Rykoff E. S. et al., 2014, ApJ, 785, 104
\bibitem[\protect\citeauthoryear{Shim et al.}{2013}]{Shim2013}
Shim H. et al., 2013, ApJS, 207, 37
\bibitem[\protect\citeauthoryear{Shogaki et al.}{2018}]{Shogaki2018}
Shogaki A. et al., 2018, in Ootsubo T., Yamamura I., Murata K., Onaka T., eds, The Cosmic Wheel and the Legacy of the AKARI Archive: From Galaxies and Stars to Planets and Life, p. 367W
\bibitem[\protect\citeauthoryear{Sunyaev $\&$ Zeldovich}{1980}]{SZ1980}
Sunyaev R. A., Zeldovich Y. B. 1980, ARA$\&$A, 18, 537
\bibitem[\protect\citeauthoryear{Stanek et al.}{2006}]{Stanek2006}
Stanek R., Evrard A. E., B\"ohringer H., Schuecker P., Nord B., 2006, ApJ, 648, 956
\bibitem[\protect\citeauthoryear{Staniszewski et al.}{2009}]{Staniszewski2009}
Staniszewski Z. et al., 2009, ApJ, 701, 32
\bibitem[\protect\citeauthoryear{Takagi et al.}{2010}]{Takagi2010}
Takagi T. et al., 2010, A$\&$A, 514, A5
\bibitem[\protect\citeauthoryear{Takey et al.}{2011}]{Takey2011}
Takey A., Schwope A., Lamer G., 2011, A$\&$A, 534, A120
\bibitem[\protect\citeauthoryear{Tamura et al.}{2016}]{Tamura2016}
Tamura N. et al., 2016, Proc. SPIE Conf. Ser. Vol. 9908, p. 99081M
\bibitem[\protect\citeauthoryear{Tempel et al.}{2014}]{Tempel2014}
Tempel E. et al., 2014, A$\&$A, 566, A1
\bibitem[\protect\citeauthoryear{Toba et al.}{2020}]{Toba2020}
Toba Y. et al., 2020, ApJ, 899, 35
\bibitem[\protect\citeauthoryear{Vulcani et al.}{2010}]{Vulcani2010}
Vulcani B. et al., 2010, ApJL, 710, L1
\bibitem[\protect\citeauthoryear{Wang et al.}{2020}]{Wang2020}
Wang T.-W. et al., 2020, MNRAS, 499, 4068
\bibitem[\protect\citeauthoryear{Wen et al.}{2012}]{Wen2012}
Wen Z. L., Han J. L., Liu F. S., 2012, ApJS, 199, 34
\bibitem[\protect\citeauthoryear{Wen $\&$ Han}{2021}]{Wen2021}
Wen Z. L., Han J. L., 2021, MNRAS, 500, 1003

\end{thebibliography}




\appendix
\section{Linking Redshift}
\label{dz}
\begin{figure*}
    \includegraphics[width=\columnwidth]{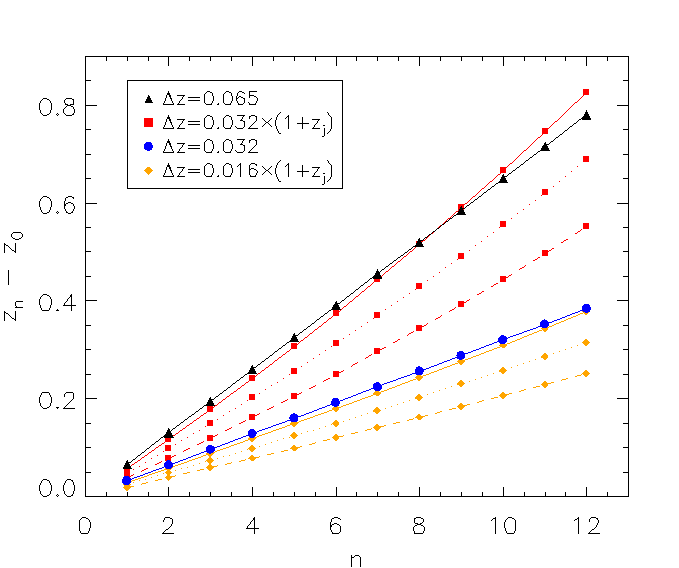}
    \includegraphics[width=\columnwidth]{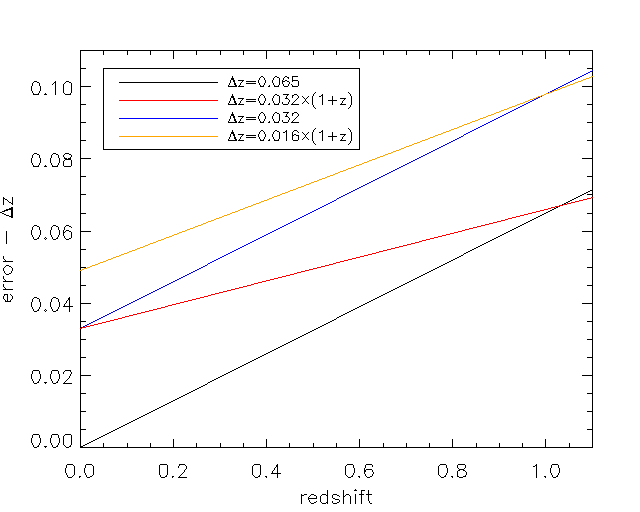}
    \caption{(left panel) The relations between the maximum redshift difference generated from linking process and the linking times in different linking redshift $\Delta z$. The dashed line, dotted line, and the solid line represent the initial redshift $z_{0}=0.2, 0.5,$ and 0.8, respectively. (right panel) The difference between photo-$z$ error, $0.065\times(1+z)$, and different $\Delta z$ as functions of redshift. }
    \label{fig:linking_z}
\end{figure*}

{
In this section, we discuss the usage of the redshift-dependent linking redshift.
\begin{equation}
   \Delta z= \Delta s \times (1+z), 
\end{equation} 
where $\Delta s$ is a constant value we choose for the linking process. Then after linking $j$ times, the redshift $z_{j}$ that the linking process can extend in the most extreme case is described in a recurrence relation: 
\begin{equation}
    z_{j}=z_{j-1}+\Delta s\times(1+z_{j-1}).     
\end{equation}
We solve this equation by adding 1 in the both sides and then making products from $j=1$ to $j=n$ to obtain 
\begin{equation}
    1+z_{n}=(1+\Delta s)^{n}(1+z_{0}).
\end{equation}
We plot the relation between the maximum redshift difference $z_{n}-z_{0}$ and the linking times $n$ in the left panel of Fig.~\ref{fig:linking_z}. We plot the relations of constant linking redshift $\Delta z=0.065$ and $\Delta z=0.032$ in black triangles and blue circles. For redshift-dependent linking redshift, the relation depends on the initial redshift $z_{0}$, and thus multiple curves of $z_{0}=0.2, 0.5,$ and 0.8 are plotted in the dashed, dotted, and solid lines, respectively. The red squares and the orange diamonds show the relations of $\Delta z=0.032\times(1+z_{j})$ and $\Delta z=0.016\times(1+z_{j})$.
We compare the constant linking redshift $\Delta z=0.032$ used in this work with the redshift-dependent linking redshift $\Delta z=0.032\times(1+z_{j})$. For the redshift-dependent one, an initial galaxy at redshift 0.5 ($z_{0}=0.5$) can be grouped with a galaxy at redshift 1.06 ($z_{10}=1.06$) after linking 10 times ($n=10$), while the linking with the constant value gives the deviation of 0.32 in this situation. Using the redshift-dependent linking redshift with a smaller $\Delta s$ can solve the widely linking problem. However, another issue arises that the difference between our photo-$z$ error, $0.065\times(1+z)$, with the linking redshift becomes larger with smaller $\Delta s$, especially at low redshift (the right panel of Fig.~\ref{fig:linking_z}). To summarise, according to this study we may not be able to claim which linking redshift is better, but at least there is no obvious benefit to using a redshift-dependent one.
}

\section{flux estimation of the Chandra cluster J175511+663354}
\label{flux_est}
\begin{figure*}
    \includegraphics[width=1.5\columnwidth]{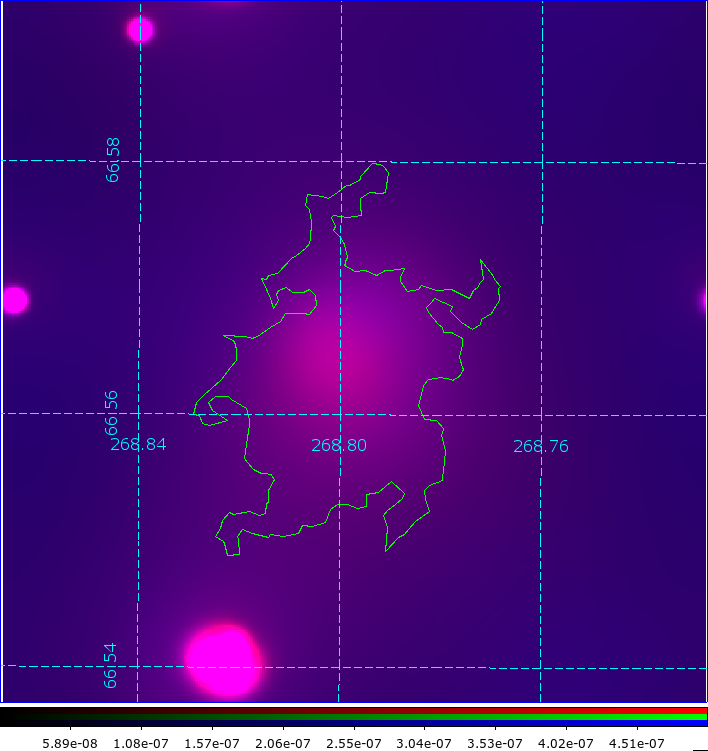}
    \caption{The smoothed Chandra image of the cluster J175511+663354. The image is stacked by images in two bands (0.5 to 2.0 keV in red and 2.0 to 8.0 keV in blue). The green polygon covers the source region detected by \textsc{vtpdetect} with the parameter \texttt{scale}=0.15. The coordinates are plotted in cyan dashed lines. The number in the colour scale bar is the count rate (in counts/sec), while the green colour in this scale bar is not used in this figure.}
    \label{fig:vtp}
\end{figure*}

We obtained the observed image \citep{Krumpe2015} around the \textit{Chandra} cluster from Stacking Analysis of \textit{Chandra} Images (CSTACK\footnote{\href{http://lambic.astrosen.unam.mx/cstack/}{http://lambic.astrosen.unam.mx/cstack/}}) developed by Takamitsu Miyaji. We detected the source of the cluster using \textsc{vtpdetect} \citep{EW1993} in \textit{Chandra} Interactive Analysis of Observations \citep[CIAO;][]{Fruscione2006}. We set the \texttt{scale} parameter of \textsc{vtpdetect} to be 0.15, so that faint sources like clusters can be detected (Fig.~\ref{fig:vtp}). The X-ray flux was estimated by Portable Interactive Multi-Mission Simulator (PIMMS\footnote{\href{https://cxc.harvard.edu/toolkit/pimms.jsp}{https://cxc.harvard.edu/toolkit/pimms.jsp}}). The input parameters are listed in Table~\ref{tab:pimms}. The estimated flux of the \textit{Chandra} cluster in 0.5 to 2.0 keV is $2.937\times10^{-14}$ erg cm$^{-2}$ s$^{-1}$.

\begin{table}
	\centering
	\caption{The input parameters for the flux estimation of the \textit{Chandra} cluster J175511+663354 using PIMMS.}
	\label{tab:pimms}
	\begin{tabular}{lr} 
		\hline
		parameter & value\\
		\hline
		Mission & CHANDRA-Cycle 12\\
		Detector & ACIS-I\\
		Grating & None\\
		Filter & None\\
		Input Energy & 0.5 to 2.0 keV\\
		Model & Plasma/APEC\\
		Galactic NH & $4\times10^{20}$ cm$^{-2}$\\
		Redshift & 0.5\\
		Redshifted NH & 0\\
		Abundance & 0.2 solar\\
		log T | keV & 7.55 | 3.0575\\
		\hline
	\end{tabular}
\end{table}

\section{author affiliations}
\textit{
$^{1}$Department of Space and Astronautical Science, Graduate University for Advanced Studies, SOKENDAI, Shonankokusaimura, Hayama, Miura District, Kanagawa 240-0193, Japan\\
$^{2}$Institute of Space and Astronautical Science, Japan Aerospace Exploration Agency, 3-1-1 Yoshinodai, Chuo-ku, Sagamihara, Kanagawa 252-5210, Japan\\
$^{3}$Institute of Astronomy, National Tsing Hua University, No. 101, Section 2, Kuang-Fu Road, Hsinchu City 30013, Taiwan\\
$^{4}$Centre for Informatics and Computation in Astronomy (CICA), National Tsing Hua University, 101, Section 2. Kuang-Fu Road, \\
$^{5}$National Astronomical Observatory of Japan, 2-21-1 Osawa, Mitaka, Tokyo 181-8588, Japan\\
$^{6}$National Institute of Technology, Wakayama College, 77 Noshima, Nada-cho, Gobo, Wakayama 644-0023, Japan\\
$^{7}$Tokyo University of Science, 1-3 Kagurazaka, Shinjuku-ku, Tokyo 162-8601, Japan\\
$^{8}$Department of Physics and Astronomy, UCLA, 475 Portola Plaza, Los Angeles, CA 90095-1547, USA\\
$^{9}$National Centre for Nuclear Research, ul. Pasteura 7, 02-093 Warsaw, Poland\\
$^{10}$Astronomical Observatory of the Jagiellonian University, ul. Orla 171, 30-244 Cracow, Poland\\
$^{11}$The Open University, Milton Keynes, MK7 6AA, UK\\
$^{12}$Department of Earth Science Education, Kyungpook National University, 80 Daehak-ro, Buk-gu, Daegu 41566, Republic of Korea\\
$^{13}$Instituto de Astronom\'ia, Universidad Nacional Aut\'onoma de M\'exico (UNAM), AP 106, Ensenada 22860, BC, Mexico\\
$^{14}$Leibniz Institut f\"ur Astrophysik Potsdam (AIP), An der Sternwarte, Potsdam, 14482, Germany\\
$^{15}$Astronomy Program, Department of Physics and Astronomy, Seoul National University, 1 Gwanak-ro, Gwanak-gu, Seoul 08826, Republic of Korea\\
$^{16}$Department of Physics and Astronomy, University College London, Gower Street, London WC1E 6BT, UK\\
$^{17}$Cosmic Dawn Center (DAWN), Copenhagen, Denmark\\
$^{18}$National Space Institute, DTU Space, Technical University of Denmark, Elektrovej 327, DK-2800 Kgs. Lyngby, Denmark\\
$^{19}$RAL Space, Rutherford Appleton Laboratory, Chilton, Didcot, Oxfordshire OX11 0QX, UK\\
$^{20}$Oxford Astrophysics, University of Oxford, Keble Rd, Oxford OX1 3RH, UK\\
$^{21}$Department of Astronomy, Kyoto University, Kitashirakawa-Oiwake-cho, Sakyo-ku, Kyoto 606-8502, Japan\\
$^{22}$Academia Sinica Institute of Astronomy and Astrophysics, 11F of Astronomy-Mathematics Building, AS/NTU, No. 1, Section 4, Roosevelt Road, Taipei 10617, Taiwan\\
$^{23}$Research Center for Space and Cosmic Evolution, Ehime University, 2-5 Bunkyo-cho, Matsuyama, Ehime 790-8577, Japan\\
$^{24}$Niels Bohr Institute, University of Copenhagen, Lyngbyvej 2, 2100 Copenhagen, Denmark\\
$^{25}$Korea Astronomy and Space Science Institute (KASI), 776 Daedeok-daero, Yuseong-gu, Daejeon 34055, Republic of Korea\\
}


\bsp	
\label{lastpage}
\end{document}